\documentclass[pdflatex,sn-aps]{sn-jnl}% American Physical Society (APS) Reference Style
\usepackage{graphicx}%
\usepackage{multirow}% 
\usepackage{amsmath,amssymb,amsfonts}%
\usepackage{amsthm}%
\usepackage{mathrsfs}%
\usepackage[title]{appendix}%
\usepackage{xcolor}%
\usepackage{textcomp}%
\usepackage{manyfoot}%
\usepackage{booktabs}%
\usepackage{algorithm}%
\usepackage{algorithmicx}%
\usepackage{algpseudocode}%
\usepackage{listings}%
\usepackage{adjustbox}
\usepackage{caption}
\usepackage{soul}
\usepackage{lipsum}
\usepackage{lineno}
\usepackage{float}
\usepackage{wrapfig} % wrap text around figures
\usepackage{txfonts}
\usepackage{subcaption} % necessary for continued figures, example in section 3 % and appendix
\usepackage{lscape} % to rotate a single page table, example in appendix.
\usepackage{placeins} % useful with \FloatBarrier, to keep 
\usepackage[nameinlink]{cleveref} %
\begin{document}

\title[Extended Universal Rotational Curve of Spiral Galaxies]{Extended Universal Rotational Curve of Spiral Galaxies}

\author*[1]{\fnm{Esha} \sur{Bhatia}}\email{b.esha@iitg.ac.in}

\author*[2,3,4]{\fnm{Paolo} \sur{Salucci}}\email{salucci@sissa.it}

\author[2,3,4]{\fnm{Tiziano} \sur{Schiavone}}\email{tiziano.schiavone@sissa.it}
\author[2,3,4]{\fnm{Sandeep} \sur{Haridasu}}\email{sandeep.haridasu@sissa.it}

% \orgdiv{Department of Physics}, 
\affil*[1]{
 \orgname{Indian Institute of Technology},
 \orgaddress{
 \city{Guwahati},
 \postcode{781039},
 \country{India}%
 }%
}

\affil[2]{% 
 \orgname{SISSA},
 \orgaddress{
 \street{Via Bonomea 265}, 
 \city{Trieste}, 
 \postcode{34136}, 
 \country{Italy}%
 }%
}
\affil[3]{
 \orgname{INFN -- Sezione di Trieste},
 \orgaddress{
 \street{Via Valerio 2},
 \city{Trieste},
 \postcode{34127},
 \country{Italy}%
 }%
}

\affil[4]{
 \orgname{IFPU},
 \orgaddress{
 \street{Via Beirut 2},
 \city{Trieste},
 \postcode{34014},
 \country{Italy}%
 }%
}


\abstract{ In this work, we aim to advance the Universal Rotation Curve (URC) paradigm by leveraging new data and extending its observational domain. Building on previous studies that established the URC using optical rotation curves reaching the galaxy optical radius $R_{opt}$, we exploit the SPARC sample's extended HI rotation curves to construct the URC out to $2\,R_{opt}$. This crucial extension enables us to investigate the mass distribution of spiral galaxies in a region dominated by dark matter—an important step to better constrain galaxy mass models and to explore the nature of the dark matter particle. We find that the URC constructed from the SPARC sample's extended HI rotation curves maintains its universal character out to $2\,R_{opt}$, with the double-normalized rotation curves collapsing onto a single profile. This extended URC provides new insights into the interplay between baryonic matter and dark matter in shaping galaxy rotation curves, particularly in the outer regions where dark matter dominates. Our results not only reinforce the URC paradigm but also refine our understanding of the mass distribution in spiral galaxies, offering new constraints on galaxy mass models and implications for the nature of dark matter.}


\maketitle
 
\section{Introduction}

Rotation Curves play a decisive role in determining the distribution of dark and luminous matter in disk galaxies. As continues to be reflected by accumulating observational results, they serve as a gateway to unraveling the nature of the dark particle. \citep{1991ApJ...368...60P} and \citep{1996MNRAS.281...27P} were the first to assert that the DM mystery at galactic scales requires investigation by means of a large sample of high-quality RCs of galaxies spanning different luminosities and Hubble types (see also the review \cite{2019A&ARv..27....2S}). This approach has revealed that disk systems exhibit a diverse family of RC profiles and amplitudes \cite{Noord}, yet all share a common distribution of {\it luminous} matter, specifically, at a first approximation, a stellar disk with a Freeman profile \citep{Freeman:1970mx}. This scenario points to a connection between the RC profiles and the structural properties of the DM halos, and counters the perspective that understanding the DM phenomenon in galaxies can be achieved by focusing on a few "representative" RCs, with the study of additional RCs only serving statistical purposes \citep{1996MNRAS.281...27P,2018FoPh,2019A&ARv..27....2S}. 

Since 1995, a substantial number of RCs extended out to the optical radius $R_{opt}$, the radius enclosing 83\% of the total luminosity of the galaxy \citep{1996MNRAS.281...27P}, have become available for Normal Spirals, Dwarf Irregulars, and Low Surface Brightness galaxies. As a result, from the rich variety in the profiles and amplitudes of individual RCs, an Universal Rotation Curve (URC) scenario has emerged: for the vast majority of disk galaxies, the circular velocity is well represented by a universal curve of the following form \citep{1996MNRAS.281...27P, Salucci:2007tm,2017MNRAS.465.4703K,2019MNRAS.490.5451D,Catinella}:
\begin{equation}\label{eq:URC}
 V_{URC}(R)=v(R/R_{opt}; \Pi_1,\Pi_2) \ V_{opt}(\Pi_1),
\end{equation}
where $v$ is a function common to all disk galaxies, $V_{opt}\equiv V(R_{opt})$, $R_{opt}$ and the parameters $\Pi_1, \Pi_2$ are observational quantities characterizing each individual galaxy.\footnote{Additional (observational) parameters can be included in the URC, if necessary, without changing its meaning.} Specifically, $R_{opt}$ (with $R_{opt} =3.2\,R_D$), in the very frequent case of a stellar Freeman disk with exponential scale-length $R_D$) represents the size of the stellar disk. $\Pi_1$ quantifies the luminous mass of the galaxy, either $V(R_{opt})$, $M_D$, or $L_{IR}$ (the stellar disk mass or its infrared luminosity), while $\Pi_2$ indicates the surface density variation between galaxies of the same "mass" $\Pi_1$, such as quantities related to: $L_{IR}/R_D^2$ or $M_D/R_D^2$. This second parameter is essential for establishing the URC in dwarf irregulars and LSB galaxies (see also the reviews \citep{2019A&ARv..27....2S,Saluccirev21,Dipaolorev,2018FoPh}). 

The URC is constructed (see: \citep{1996MNRAS.281...27P}) by co-adding many individual RCs with good spatial resolution, extended to radii comparable to the edge of the stellar disk $R_{opt}$, roughly $1/20$ of $R_{vir}$, the size of the dark halo \citep{Mo:1997vb, Kravtsov2012THESR}. The significant diversity observed in RCs, when they are expressed in physical units $(V(R), R)$ in $\mathrm{(km/s,\,kpc)}$ (see \cref{fig:urc_al}), mostly disappears when, (a) we adopt {\it double normalized} radial and velocity coordinates as in \cref{eq:VURC} (see \cref{fig:urc_norm}):

\begin{equation}
 (V(R),R) \rightarrow \Bigg(\frac {V\big(\frac{R}{ R_{opt}}\big)}{V(R_{opt})}, \frac {R}{R_{opt}}\Bigg)
 \label{eq:VURC} 
 \end{equation}
\vskip 0.5cm
and (b) we account for the dependence of the individual RCs on the two observable physical quantities as described in \cref{eq:URC}. Careful analysis of large samples of RCs has shown the existence of a Universal Curve, such as that in \cref{eq:URC}, representing the great majority of the RCs disk systems of the local Universe \citep{2017MNRAS.465.4703K, 1991ApJ...368...60P,2019MNRAS.490.5451D, 1996MNRAS.281...27P,Lopez}. For the issue of the URC in disk systems see also \citep{LSD,2007MNRAS.377..507Y} and the reviews \citep{2019A&ARv..27....2S,
Saluccirev21,Dipaolorev, 2018FoPh}. 
% and in diverse implementations \cite{Rhee, Lopez,Catinella, Noord} and 

Complementary studies support this phenomenological view. Principal-component analyses indicate that the diversity of spiral-galaxy rotation curves can be captured by a small number of dominant modes \citep{Rhee}, while template RCs from large homogeneous samples reveal regular average trends across luminosity classes out to several disk scale-lengths \citep{Catinella}. Investigations of early-type disks likewise show that RC shape correlates with optical properties, including the occurrence of declining outer curves in some systems \citep{Noord}. More recently, purely empirical velocity-profile parametrizations have shown that large RC datasets can be fit accurately without adopting {\it a priori} a fixed decomposition into luminous and dark components \citep{Lopez}.

% See also \cite{Rhee, Lopez,Catinella, Noord}, 

It is worth noting that, in disk systems, exploring the mass structure of galaxies through a) $\sim 100$ individual high-quality RCs and b) the Universal Curve derived from coaddition of $\sim 1000$ good quality RCs are complementary strategies that arrive to an unique scenario. The {\it individual} RCs yield the density distribution of DM halos, while the URC reveals relationships among the various dark and luminous structural parameters. In the recent works the URC is built observationally out to $R_{opt}$ using RC data, and then it is extrapolated to $R_{vir}$ through suitable galaxy mass models constrained by the available kinematics inside $R_{opt}$ \citep{Salucci:2007tm, 2017MNRAS.465.4703K,2019MNRAS.490.5451D}. 

Let us stress that the mass modeling of {\it individual} high quality RCs of different luminosity shows that (see e.g. \cite{2019A&ARv..27....2S}) the structural parameters of the galaxy's dark and luminous components are related in a way that implies the existence of the (claimed) URC scenario. On the other hand, the mass modeling of the {\it coadded} URC reveals, even more clearly, the existence of the same scaling laws. Remarkably, the processes of co-adding and suitably averaging the individual RCs take care of their observational errors and their kinematical disturbances unrelated to the galaxy mass distribution, so that the intrinsic systematics of the latter can clearly emerge.

Given the growing importance of galaxy kinematics in revealing the particle nature of dark matter, and the recent availability of very extended HI RCs, it is timely to construct the URC out to $2\,R_{opt}$ using individual RCs, thereby probing a region where the DM component dominates the gravitational potential in all objects.

In this work, we utilize the well-known SPARC sample \citep{2016AJ....152..157L}, selecting extended, high-resolution, high-quality HI RCs to construct the URC in the region $3.2 - 6.0\,R_D$, thus doubling the radial domain over which the URC is observationally established using optical RCs. Notably, the HI RCs from this sample are found to combine well with high-resolution $H\alpha$ RCs in the inner regions and display minimal beam smearing, ensuring well-resolved kinematics across each galaxy. To this end, we (i) select a subsample of high-quality SPARC HI rotation curves with adequate radial coverage, (ii) construct double-normalized coadded rotation curves in velocity and radial bins, and (iii) derive an empirical URC representation over $R_{opt}\le R\le 2R_{opt}$. We also perform an independent coaddition of individual RCs around $R_{opt}$ using the PROBES sample and compare the resulting URC with the URC-SPARC and the URC-vir. 

Finally, we investigate, with the SPARC sample, the outer galaxy regions by establishing there the Radial Tully-Fisher relation, a family of relationships between the galaxy magnitudes and the corresponding circular velocities at fixed $R/R_{opt}$ radii.

This work is structured as follows: in \Cref{sec:sparc_cat}, we describe the SPARC sample and the selection criteria for the RCs used in our analysis. In \Cref{sec:urc2p0}, we present the construction of the URC out to $2\,R_{opt}$, discuss its properties, and compare it with previous URC determinations. {In \Cref{sec:RTF}, we investigate the galaxy luminosity and kinematic properties for the SPARC sample through the Radial Tully-Fisher (RTF) relation.} Finally, in \Cref{sec:conc}, we summarize our findings and discuss their implications for galaxy mass models and dark matter research.

\section{The URC from the SPARC Sample}
\subsection{SPARC catalog}
\label{sec:sparc_cat}

The Spitzer Photometry and Accurate Rotation Curve (SPARC) catalog comprises $175$ rotationally supported galaxies, spanning the morphological range from spirals to irregulars \citep{2016AJ....152..157L}\footnote{Publicly available at \href{https://astroweb.case.edu/SPARC/}{https://astroweb.case.edu/SPARC/}.}. The catalogue provides $\mathrm{HI}$ and $H\alpha$ rotation curves \footnote{For the purposes of this work, only the HI rotation curves are utilized.} for disk galaxies, the majority of which are located within $\sim 10\,\mathrm{Mpc}$ of the Milky Way\footnote{Distances are determined through various methods, including the Tip of the Red Giant Branch (TRGB) \citep{2018ApJ...858...62K, 2019ApJ...880...63M}, Cepheid period-luminosity calibration \citep{Fouque:2003tm}, Type Ia Supernovae, and the Hubble flow.}. In this sample, the HI rotation curves extend well beyond the optical radius with excellent spatial resolution. Specifically, the $21\,\mathrm{cm}$ line data are obtained from multiple radio interferometers, including the Very Large Array, Westerbork Synthesis Radio Telescope \citep{1996ApJS..107..239T}, Australian Telescope Compact Array \citep{2013MNRAS.432.1294P}, and the Giant Metrewave Radio Telescope; these feature beam sizes ranging from $B=13''$ to $B=45''$ \citep{2001AJ....122.2381M, 2009A&A...493..871S}, corresponding to approximately $300\ B/30^"\, D/(2\,\mathrm{Mpc})\ \mathrm{pc}$\footnote{This restricts the number of galaxies whose kinematics can be traced via HI data.}. For the stellar component, the catalogue provides fluxes and magnitudes in the Spitzer $3.6\,\mu$m band, an excellent estimator of the spiral disks stellar masses \cite{Meidt:2014mqa}, spanning from $10^7\text{ to }10^{12}\,L_{\odot,3.6}$, along with the scale length of the exponential Freeman disk $R_D$, the HI disk mass, and its scale length. 

%\cite{McGaugh:2016leg, Rodrigues:2017vto, Salucci:2018hqu, deAlmeida:2018kwq, Li:2020iib, Khelashvili:2024gus}
The individual rotation curves (RCs) in the SPARC sample enable detailed mass modeling for a substantial number of galaxies, providing insight into both their luminous and dark matter components (e.g. \citep{2016PhRvL.117t1101M, 2017MNRAS.470.2410R, 2018JCAP...08..012D, 2020ApJS..247...31L,2024arXiv240110202K}.) In this work, we leverage the high-quality HI kinematics from SPARC to significantly extend the observational domain of the Universal Rotation Curve: specifically, we double the region over which the URC is directly constructed, increasing its maximum radius from $3.2\,R_D$ to $6\,R_D$. This expanded radial range allows us to (a) test directly whether the URC holds at larger radii, (b) construct the URC in this outer regime, and (c) compare it to the URC extrapolated from the inner region ($R = 0$--$3.2\,R_D$), previously analyzed using optical rotation curves. Extending the URC out to $2 R_{opt}$ not only enlarge the region where it represents a fundamental property of galaxy dynamics but also substantially enhances our understanding of spiral galaxy mass models. In this outer region, the declining contribution of the stellar disk to the circular velocity, in a Keplerian way, makes it significantly easier to disentangle the dark and luminous matter components, with the dark matter halo becoming dominant in all systems.

{\renewcommand{\arraystretch}{1.2} 
\setlength{\tabcolsep}{6pt} 

\begin{table*}[t!]
\caption{Details of the eight $\langle V_{opt}\rangle$ bins of the present analysis. Column I: The average optical velocity $\langle V_{opt}\rangle$ value in each bin. Columns II and III: the maximum and minimum values of $V_{opt}$ among the galaxies in the bin. Column IV shows the number of the galaxies in each bin. }
\vskip 0.5truecm
\label{tab:bin_opticalval}
\centering
\begin{tabular}{cccc}
\toprule
$\langle V_{{opt}}\rangle$(km/s) & $V_{opt, \rm{min}}$(km/s) & $V_{opt, \rm{max}}$(km/s) & $N_{\rm gal}$\\ 
\midrule
$42.9$&$20.8$&$57.4$&$15$ \\

$68.7$&$60.5$&$76.8$&$18$\\

$81.8$&$77.4$&$90.4$&$9$\\

$112.5$&$103.9$&$126.3$&$14$\\

$147.5$&$129.9$&$161.5$&$7$\\

$190.0$&$169.6$&$209.5$&16\\

$226.1$&$210.0$&$245.6$&$15$\\

$295.5$&$254.5$&$375.2$&$11$\\
\botrule
\end{tabular}

\end{table*}
}

{\renewcommand{\arraystretch}{1.2} 
\setlength{\tabcolsep}{6pt} 
\begin{table*}[t!]
\caption{The coordinates of the $6$ radial bins adopted for the $105$ qualified SPARC galaxies. First column indicates the bin number, the second and the third its innermost and outermost positions: $r=R/R_{{opt}}$. }
\label{tab:bin_rn}
\centering
\begin{tabular}{@{}lll@{}}
\toprule
Bin & $r_{min}$ & $r_{max}$ \\ 
\midrule
$1$ & $1.000$ & $1.099$\\

$2$ & $1.100$ & $1.199$\\

$3$ & $1.201$ & $1.399$\\

$4$ & $1.402$ & $1.599$\\

$5$ & $1.600$ & $1.800$\\

$6$ & $1.803$ & $1.872$\\
\botrule

\end{tabular}

\end{table*}
}

Among the $175$ galaxies of SPARC, $22$ galaxies, as in the original analysis, are rejected because either they are face-on galaxies or the overall quality of the RC is very low, see
\citep{2016AJ....152..157L}. By following \citep{1996MNRAS.281...27P}, we begin by determining the optical radius values ($R_{opt}$) for each galaxy in the sample from the auxiliary data given that $ R_{opt}\equiv 3.2 \ R_D$. From the individual rotation curves (RCs), we determine $V_{opt} \equiv V(R_{opt})$ with a precision of a few percent. Subsequently, (a) each SPARC RC is assigned to one of eight velocity bins (hereafter, {\bf VEL} bins) based on its $V_{opt}$ value, such that the entire SPARC sample is distributed across these bins. \cref{tab:bin_opticalval} summarizes, for each {\bf VEL} bin, the mean $\langle V_{opt}\rangle$, the minimum and maximum $V_{opt}$, and the number of galaxies included. Then, (b) within each {\bf VEL} bin, the RC measurements of member galaxies are further grouped into six radial bins (hereafter, {\bf RAD} bins), the boundaries of which ($r_{min}, r_{max}$, with $r = R/R_{opt}$) are provided in \cref{tab:bin_rn}.

In constructing the URC, we implemented two additional selection criteria to ensure the suitability of the RCs included in our analysis. First, we consider only those (SPARC) RCs that exhibit no significant non-circular motions and that provide adequate sampling within the region of interest. Specifically, an RC is included if it contributes data points to either: {\it (a)} at least three distinct {\bf RAD} bins, or {\it (b)} two {\bf RAD} bins, provided that the outermost measurement falls within the 4th or 5th {\bf RAD} bin. These requirements guarantee that each accepted RC offers sufficient spatial coverage to robustly construct the coadded RCs in the relevant region. As a result of these criteria, we exclude 48 RCs\footnote{This includes RCs with fewer than two measurements across the six bins, or cases where both measurements are restricted to bins 1–3.} which, despite their good quality inside $R_{opt}$, lack the necessary measurements in the region under investigation or display pronounced non-axi-symmetric motions in that outer region. Consequently, our final sample comprises $105$ SPARC RCs used in the present analysis.

\begin{figure*}[h]
 {\centering
\includegraphics[width=11.5cm]{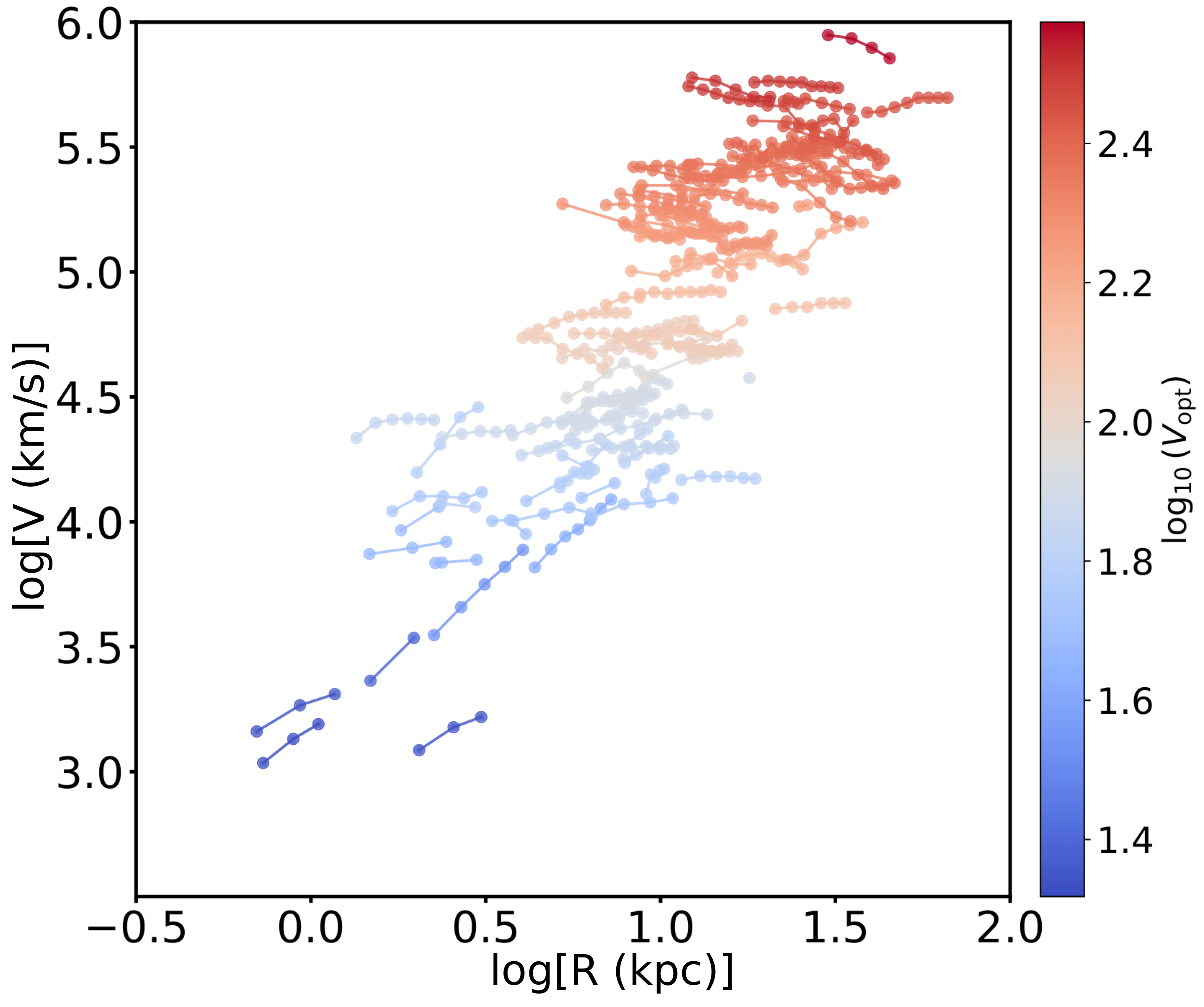}
\par}
 \caption{ The circular velocities as a function of galactocentric radius for the final sample of 105 SPARC rotation curves, plotted over the range: $3.2\,R_D$--$6\,R_D$.  The RC profiles have a low statistical correlation with $V_{opt}$ data from \citep{2016AJ....152..157L}.} 
 \label{fig:urc_al}
\end{figure*}

\begin{figure*}[h!]
\centering
\includegraphics[width=11.5cm]{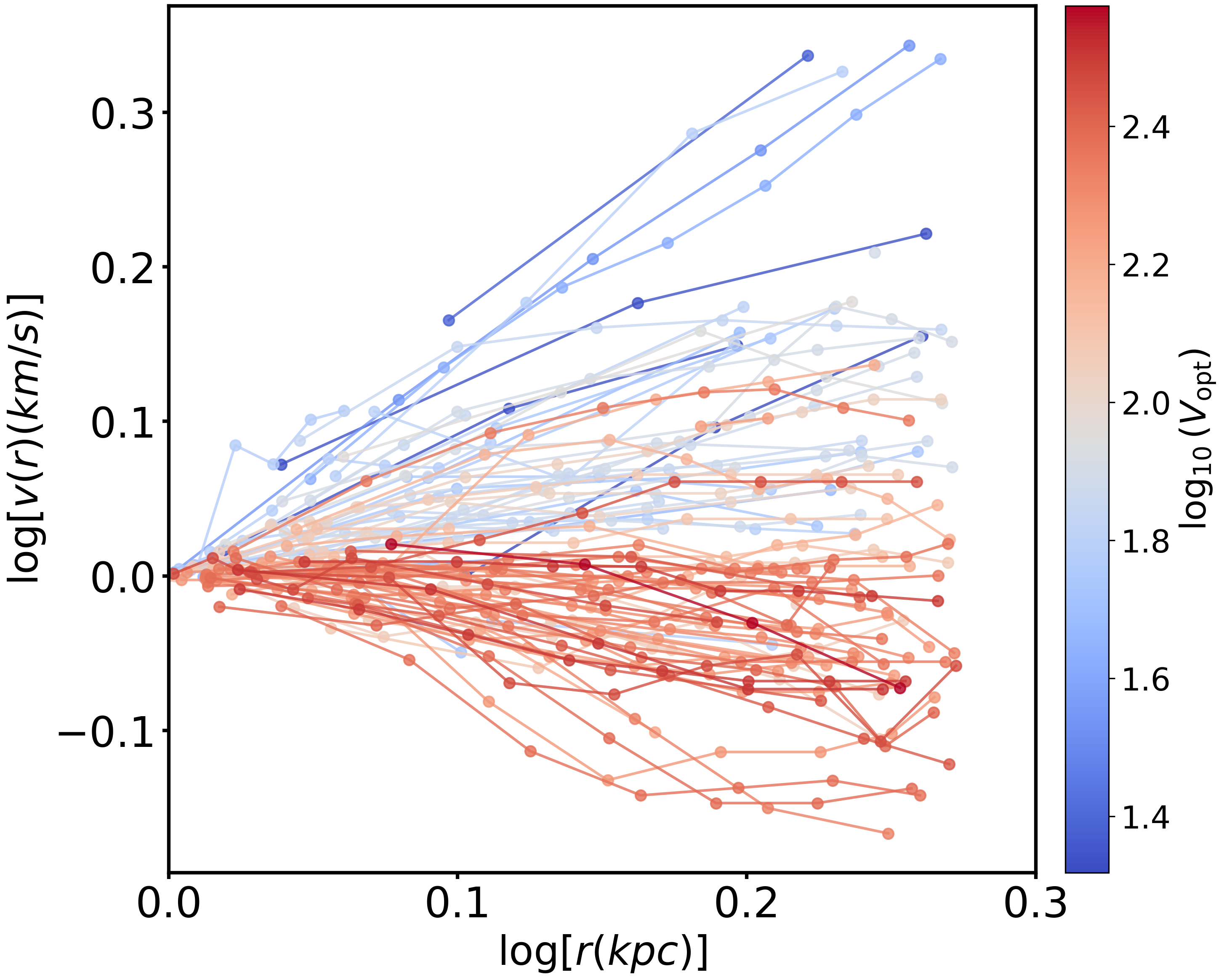}
 \caption{Double-normalized extended SPARC rotation curves (RCs) as shown in \cref{fig:urc_al}, but with $r = R / R_{opt}$ and $v(r) = V(R) / V_{opt}$. In contrast to \cref{fig:urc_al}, all RCs in the sample collapse into a single family, with $V_{opt}$ acting as a labeling parameter that identifies each curve.}
 \label{fig:urc_norm}
\end{figure*}

\Cref{fig:urc_al} shows the observed circular velocities for the 105 SPARC galaxies that meet the selection criteria described above. In \cref{fig:urc_norm}, we present the same rotation curves, but with both the galactocentric radii ($R$) and velocities $V(R)$ double-normalized (hereafter $\bf 2N$) using their respective $R_{opt}$ and $V_{opt}$ values. It is remarkable that such a double-normalization, indicated by \cref{eq:VURC}, transforms a chaotic ensemble of $V(R)$ functions into an ordered ensemble of $v(r)$ functions, exactly as it was observed in previous constructions of the URC \citep{2019MNRAS.490.5451D, 1996MNRAS.281...27P, 2017MNRAS.465.4703K}. This transformation reveals the underlying universality in the rotation curves of spiral and dwarf irregular galaxies, suggesting that their kinematics can be described by a common functional form when appropriately scaled by the value at the optical radius of specific quantities. Accordingly, the data presented in \cref{fig:urc_norm} serve as the foundation for constructing the URC-2opt from SPARC sample using the established methodology \citep{ 1996MNRAS.281...27P, Salucci:2007tm, 2017MNRAS.465.4703K}. Please see \Cref{sec:urc2p0} for details on the construction process.
 
{\renewcommand{\arraystretch}{1.6} 
\setlength{\tabcolsep}{6pt} 
\begin{table}[h!]
\caption{Compilation of the URC studies. Column 1: acronym of the work, Column 2: number of the sampled galaxies, Column 3: the Hubble type, Column 4: the region probed for URC: $R_{opt}$ and $R_v$, the optical and virial radius, respectively, Column 5: Reference.}
\label{tab:cat_all}
\centering
% \begin{adjustbox}{angle=90}
\begin{tabular}{cccccc}
\toprule
$Name$ & Number of Galaxies & Hubble type& Extension & Reference & Notes\\ 
(1) & (2) & (3) & (4) & (5) &(6) \\
\midrule
\midrule
URC-0 & 60 & Late-spirals & $R_{opt}$ & \citep{1991ApJ...368...60P} & \footnotemark[6]\\

URC-opt & $1100$ & Late-spirals & $R_{opt}$ & \citep{1996MNRAS.281...27P} & \footnotemark[7] \\

URC-vir & $1100$ & Late-spirals & $R_{opt}$\footnotemark[1], $R_v$\footnotemark[2] & \citep{Salucci:2007tm} & \footnotemark[8]\\

URC-dIrr& $36$ & dwarf-Irregulars & $R_{opt}$\footnotemark[1], $R_v$\footnotemark[2] & \citep{2017MNRAS.465.4703K} & \footnotemark[9]\\

URC-LSB & $73$ & LSB & $R_{opt}$\footnotemark[1], $R_v$\footnotemark[2]& \citep{2019MNRAS.490.5451D} & \footnotemark[10] \\

% URC2.0 & $105$ & S+dIrr & $R_{opt} - 2R_{opt}$ & \begin{tabular}{@{}c@{}}This work \\ Data: \citep{2016AJ....152..157L}\end{tabular} \\
URC-2opt & $105$ & S+dIrr & $R_{opt} - 2R_{opt}$ & This work\footnotemark[3] & \footnotemark[11] \\ 
URC-zoom & $1118$ & {Varied}\footnotemark[5] & $0.95\, R_{opt} - 1.2 \,R_{opt}$ & This work\footnotemark[4] & \footnotemark[12]\\

\botrule
\end{tabular} 
% \end{adjustbox}
\footnotetext[1]{URC built from observations out to $R_{opt}$.} 
\footnotetext[2]{URC built from observations out to $R_{opt}$ and extrapolated out to $R_{v}$ by means of appropriate mass models.} 
\footnotetext[3]{Utilizing the SPARC dataset \citep{2016AJ....152..157L}.}
\footnotetext[4]{Utilizing the PROBES compendium \citep{2022ApJS..262...33S}.}

\footnotemark[5]{The sample of galaxies in the PROBES compendium is varied in terms of Hubble type, luminosity, and mass. A detailed analysis based on the Hubble type is deferred to a future work.}

\footnotemark[6]{Original claim of the URC.}\\
\footnotemark[7]{URC for Spirals, out to $R_{opt}$}\\
\footnotemark[8]{URC for Spirals, out to the virial radius.}\\
\footnotemark[9]{URC for dwarf Irregulars.}\\
\footnotemark[10]{URC for Low Surface Brightness galaxies.}\\
\footnotemark[11]{URC from the SPARC sample.}\\
\footnotemark[12]{Zoomed URC from the PROBES compendium.}
\end{table}
}

\subsection{URC-2opt from SPARC}
\label{sec:urc2p0}

The URC in spiral, dwarf irregulars, and LSB galaxies \citep{Salucci:2007tm,2019MNRAS.490.5451D,2017MNRAS.465.4703K} is built from direct RC measurements taken at radii out to (1-1.2) $R_{opt}$ and from a proper extrapolation of the latter out to $R_{vir}$ (see \cref{tab:cat_all} for previous URC determinations). 

It is important to clarify that we do not utilize the SPARC sample to derive the URC in the inner region ($0 < R \leq R_{opt}$). This domain has already been thoroughly explored using extensive samples of $H_\alpha$/Optical RCs—datasets that notably include the majority of the SPARC RCs themselves. For studies within this inner regime, $H_\alpha$ RCs are preferable due to their notably higher spatial resolution compared to HI data. Additionally, the SPARC sample encompasses S, dIrr, and LSB galaxies, which exhibit some diversity in their inner URCs. In contrast, in the outer region investigated here ($R_{opt} \leq R \leq 2\, R_{opt}$), the context changes significantly: $H_\alpha$ emitters are rare or absent, and the rotation curve profiles in this range display a much weaker dependence on Hubble type and kinematic spatial resolution. Consequently, employing SPARC HI RCs to investigate the URC in this outer region is both appropriate and necessary.

{\renewcommand{\arraystretch}{1.2} 
\setlength{\tabcolsep}{6pt} 
\begin{table*}[t!]
 \caption{The table lists the coadded RCs, obtained by combining $6$ {\bf RAD} bins within each of the $8$ optical {\bf VEL} bins of the SPARC sample. The index $i$ denotes the {\bf RAD} bin, $N$ is the number of measurements in each bin, $r_i$ is the mean normalized radius, $v_i$ is the mean normalized velocity, and $\sigma_i$ is the standard deviation of the latter quantity. We also report the quantities $R_i$ and $V_i$ in physical units ($\textrm{kpc}$ and $\textrm{km s}^{-1}$, respectively).}
 \label{tab:avg-vopt}
 \centering
 \begin{minipage}{0.48\textwidth}
 \centering
 %\caption{Average $V_{opt}=44.77\,\mathrm{km\,s^{-1}}$}
 \small
 \resizebox{\textwidth}{!}{%
 \begin{tabular}{@{}r r r r r r r r@{}}
 \hline
 \multicolumn{7}{c}{$\langle V_{opt} \rangle = 42.9\,\mathrm{km\,s^{-1}}$} \\
 \hline
 $i$ & $N$ & $r_i$ & $v_i$ & $\sigma_i$& $R_i$ & $V_i$\\
 \hline
 $1$ & $10$ & $1.031$ & $1.012$ & $0.007$ & $2.66$ & $43.37$\\
 $2$ & $~3~$ & $1.129$ & $1.035$ & $0.017$& $2.92$ & $44.39$ \\
 $3$ & $14$ & $1.274$ & $1.077$ & $0.019$ & $3.29$ & $46.19$ \\
 $4$ & $~10~$ & $1.481$&$1.139$ &$0.022$ & $3.83$ & $48.86$ \\
 $5$ & $~10~$ & $1.664$ & $1.194$	&$0.043$ & $4.31$ & $51.19$ \\
 $6$ & $~5~$ & $1.824$ & $1.261$& $0.064$ & $4.72$ & $54.09$ \\
 \hline
 \end{tabular}
 }
 \resizebox{\textwidth}{!}{%
 \begin{tabular}{@{}r r r r r r r r@{}}
 \hline
 \multicolumn{7}{c}{$\langle V_{opt} \rangle = 81.8\,\mathrm{km\,s^{-1}}$} \\
 \hline
 $i$ & $N$ & $r_i$ & $v_i$ & $\sigma_i$& $R_i$ & $V_i$\\
 \hline
 $1$ & $~3~$ & $1.079$ & $1.033$ & $0.009$ & $6.52$ & $84.51$\\
 $2$ & $~4~$ & $1.141$ & $1.039$ & $0.017$& $7.14$ & $85.07$ \\
 $3$ & $10$ & $1.300$ & $1.076$ & $0.012$ & $8.05$ & $88.07$ \\
 $4$ & $11$ & $1.512$&$1.095$ &$0.013$ & $9.36$ & $89.57$ \\
 $5$ & $11$ & $1.712$ & $1.153$	&$0.014$ & $10.52$ & $94.34$ \\
 $6$ & $~5~$ & $1.843$ & $1.135$& $0.018$ & $11.53$ & $92.91$ \\
 \hline
 \end{tabular}
 }
 \resizebox{\textwidth}{!}{%
 \begin{tabular}{@{}r r r r r r r r@{}}
 \hline
 \multicolumn{7}{c}{$\langle V_{opt} \rangle = 147.5\,\mathrm{km\,s^{-1}}$} \\
 \hline
 $i$ & $N$ & $r_i$ & $v_i$ & $\sigma_i$& $R_i$ & $V_i$\\
 \hline
 $1$ & $4$ & $1.047$ & $1.003$ & $0.007$ & $11.62$ & $147.94$\\
 $2$ & $7$ & $1.144$ & $1.014$ & $0.006$& $12.74$ & $149.52$ \\
 $3$ & $8$ & $1.317$ & $1.026$ & $0.016$ & $14.37$ & $151.23$ \\
 $4$ & $8$ & $1.499$&$1.028$ &$0.024$ & $16.70$ & $151.64$ \\
 $5$ & $9$ & $1.689$ & $1.059$	&$0.016$ & $18.77$ & $156.20$ \\
 $6$ & $3$ & $1.837$ & $1.026$& $0.012$ & $20.57$ & $151.25$ \\
 \hline
 \end{tabular}
 }
 \resizebox{\textwidth}{!}{%
 \begin{tabular}{@{}r r r r r r r r@{}}
 \hline
 \multicolumn{7}{c}{$\langle V_{opt} \rangle = 226.1\,\mathrm{km\,s^{-1}}$} \\
 \hline
 $i$ & $N$ & $r_i$ & $v_i$ & $\sigma_i$& $R_i$ & $V_i$\\
 \hline
 $1$ & $12$ & $1.050$ & $1.001$ & $0.003$ & $16.83$ & $226.35$\\
 $2$ & $13$ & $1.154$ & $1.004$ & $0.005$& $18.44$ & $226.93$ \\
 $3$ & $15$ & $1.306$ & $0.987$ & $0.011$ & $20.80$ & $223.13$ \\
 $4$ & $25$ & $1.496$&$0.980$ &$0.013$ & $24.18$ & $221.74$ \\
 $5$ & $23$ & $1.686$ & $0.976$	&$0.014$ & $27.17$ & $220.76$ \\
 $6$ & $~7~$ & $1.838$ & $0.966$& $0.032$ & $29.78$ & $218.49$ \\
 \hline
 \end{tabular}
 }
 \end{minipage}
 \hfill
 \begin{minipage}{0.48\textwidth}
 \centering
 %\caption{Average $V_{opt}=69.32\,\mathrm{km\,s^{-1}}$}
 
 \small
 \resizebox{\textwidth}{!}{%
 \begin{tabular}{@{}r r r r r r r r@{}}
 \hline
 \multicolumn{7}{c}{$\langle V_{opt} \rangle = 68.7\,\mathrm{km\,s^{-1}}$} \\
 \hline
 $i$ & $N$ & $r_i$ & $v_i$ & $\sigma_i$& $R_i$ & $V_i$\\
 \hline
 $1$ & $10$ & $1.050$ & $1.030$ & $0.010$ & $5.34$ & $70.71$\\
 $2$ & $15$ & $1.142$ & $1.058$ & $0.010$& $5.85$ & $72.63$ \\
 $3$ & $22$ & $1.288$ & $1.061$ & $0.010$ & $6.60$ & $72.87$ \\
 $4$ & $16$ & $1.505$&$1.101$ &$0.021$ & $7.67$ & $75.58$ \\
 $5$ & $~8~$ & $1.699$ & $1.109$	&$0.046$ & $8.62$ & $76.16$ \\
 $6$ & $~3~$ & $1.832$ & $1.134$& $0.024$ & $9.45$ & $77.85$ \\
 \hline
 \end{tabular}%
 }
 \resizebox{\textwidth}{!}{%
 \begin{tabular}{@{}r r r r r r r r@{}}
 \hline
 \multicolumn{7}{c}{$\langle V_{opt} \rangle = 112.5\,\mathrm{km\,s^{-1}}$} \\
 \hline
 $i$ & $N$ & $r_i$ & $v_i$ & $\sigma_i$& $R_i$ & $V_i$\\
 \hline
 $1$ & $10$ & $1.058$ & $1.008$ & $0.003$ & $8.67$ & $113.38$\\
 $2$ & $17$ & $1.138$ & $1.011$ & $0.007$& $9.51$ & $113.86$ \\
 $3$ & $21$ & $1.297$ & $1.009$ & $0.009$ & $10.72$ & $113.51$ \\
 $4$ & $18$ & $1.504$&$1.014$ &$0.012$ & $12.46$ & $114.10$ \\
 $5$ & $17$ & $1.710$ & $1.013$	&$0.016$ & $14.01$ & $113.94$ \\
 $6$ & $~2~$ & $1.855$ & $1.065$& $0.056$ & $15.35$ & $119.81$ \\
 \hline
 \end{tabular}
 }
 \resizebox{\textwidth}{!}{%
 \begin{tabular}{@{}r r r r r r r r@{}}
 \hline
 \multicolumn{7}{c}{$\langle V_{opt} \rangle = 190.0\,\mathrm{km\,s^{-1}}$} \\
 \hline
 $i$ & $N$ & $r_i$ & $v_i$ & $\sigma_i$& $R_i$ & $V_i$\\
 \hline
 $1$ & $11$ & $1.048$ & $1.000$ & $0.002$ & $~9.40~$ & $190.20$\\
 $2$ & $~9~$ & $1.156$ & $0.997$ & $0.004$& $10.30$ & $189.39$ \\
 $3$ & $17$ & $1.303$ & $0.978$ & $0.005$ & $11.62$ & $185.92$ \\
 $4$ & $19$ & $1.489$&$0.963$ &$0.011$ & $13.51$ & $183.02$ \\
 $5$ & $21$ & $1.695$ & $0.959$	&$0.009$ & $15.18$ & $182.27$ \\
 $6$ & $~3~$ & $1.825$ & $0.943$& $0.010$ & $16.64$ & $179.17$ \\
 \hline
 \end{tabular}
 }
 \resizebox{\textwidth}{!}{%
 \begin{tabular}{@{}r r r r r r r r@{}}
 \hline
 \multicolumn{7}{c}{$\langle V_{opt} \rangle = 295.5\,\mathrm{km\,s^{-1}}$} \\
 \hline
 $i$ & $N$ & $r_i$ & $v_i$ & $\sigma_i$& $R_i$ & $V_i$\\
 \hline
 $1$ & $~8~$ & $1.050$ & $0.999$ & $0.003$ & $21.05$ & $295.23$\\
 $2$ & $11$ & $1.156$ & $0.997$ & $0.005$& $23.07$ & $294.69$ \\
 $3$ & $13$ & $1.314$ & $0.987$ & $0.009$ & $26.02$ & $291.67$ \\
 $4$ & $17$ & $1.500$&$0.964$ &$0.009$ & $30.25$ & $284.99$ \\
 $5$ & $16$ & $1.713$ & $0.952$	&$0.013$ & $33.99$ & $281.41$ \\
 $6$ & $~5~$ & $1.847$ & $0.958$& $0.031$ & $37.26$ & $283.19$ \\
 \hline
 \end{tabular}
 }

 \end{minipage}

\end{table*}
}

Following \citep{1996MNRAS.281...27P}, the first step in generating the URC is the double-normalization ${\bf 2N}$ (defined in Section~\ref{sec:sparc_cat}) of the available rotation velocities $\rm{V(R)}$, according to which, in each RC, the radial coordinate $R$ is scaled/normalized by the value of optical radius $R_{opt}=3.2 R_D$ and, similarly, the rotation velocity $V(R/R_{opt})$ is normalized by $V_{opt}$, its value at the optical radius Then: $ v(r)\equiv V(R/R_{opt}) /V_{opt}$, $ r\equiv R/R_{opt}$\footnote{To transform the input data $V(R)$ into $ V(R/R_{opt}) /V_{opt}$, helps in investigating the galactic {\it distribution} of the (dark+luminous) matter.}. In detail, the 105 ${\bf 2N}$ RCs are grouped into $8$ $\bf VEL$ bins according to their $V_{opt} $ values. For each bin, the number $N$ of its RCs, its limiting $V_{opt} $ values and its average $\langle V_{opt}\rangle $ value are indicated in \cref{tab:bin_opticalval}. The whole SPARC sample ranges from $\langle V_{opt}\rangle\sim 43\, \rm km/s $ to $\langle V_{opt}\rangle=295 \ \rm km/s $ (see (\ref{tab:avg-vopt})) and covers both the spirals and dwarf irregulars families. 

\begin{figure*}
 \centering
 \includegraphics[width=\linewidth]{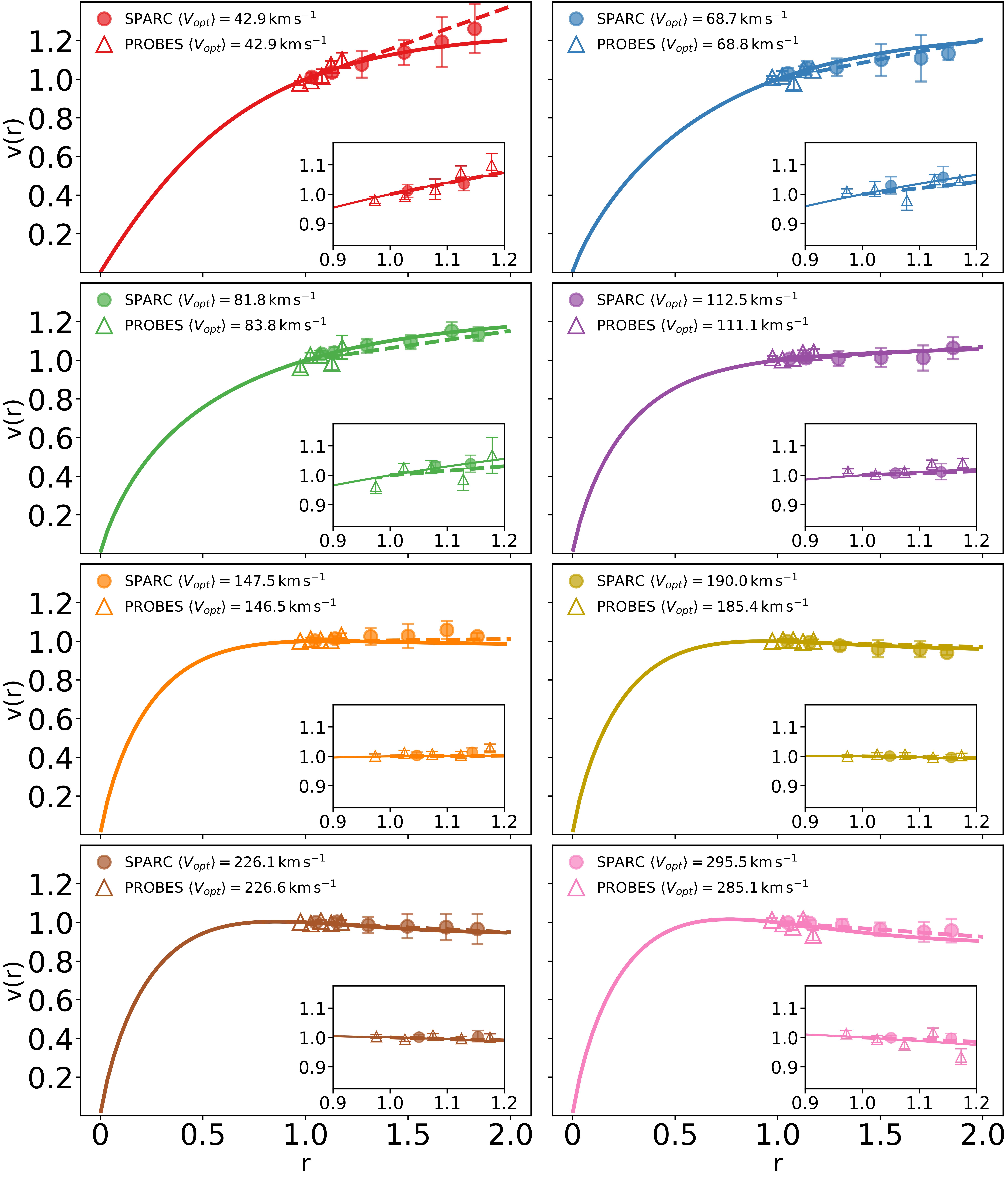}
 \caption{The URC-2opt from SPARC data in $8$ $\langle V_{opt}\rangle$ bins. The 6 values of this URC built in the region $3\rm{R_D}$ to $6\rm{R_D}$ are represented by filled circles. The solid line corresponds to the URC-vir predictions \citep{Salucci:2007tm} and the dashed line is \cref{eq:vurc}. The open triangle represents URC-zoom, the URC evaluated from the PROBES Sample in the region ($0.95~R_{opt}$, $1.20~R_{opt}$). In the inset of the panels, the latter region is zoomed for a better comparison of the two URCs. }
 \label{fig:binVopt}
\end{figure*} 

For each of the 8 $\bf VEL$ bin, the ${\bf 2N}$ RC data are arranged in 6 $\bf RAD$ bins, defined in \cref{tab:bin_rn}. This involves averaging in 48 $\bf RAD$ bins to obtain 8 coadded ${\bf 2N}$ RCs (see \cref{tab:avg-vopt} and \cref{fig:binVopt}) that represent the SPARC kinematics. 
Let us stress that, in principle, a set of coadded ${\bf 2N}$ RCs as the above one, can be constructed out of any data sample, independently of whether the coaddition process yields to an Universal curve with a physical justified profile. In fact, such coadded curves are observational quantities very effective for testing the outcome of simulations and/or of theoretical predictions of specific cosmological scenarios. 
However, here, as in the previous works reported in \cref{tab:cat_all}, relying on the SPARC data, we are more ambitious. With no information from the RCs inside $R_{opt}$, we want to probe the existence and then build the URC-2opt, \textit{i.e.}, the URC in the region $R_{opt} \leq R\leq 2\ R_{opt}$ of Spiral and dwarf Irregular galaxies. In other words, we build a number of {\bf 2N } coadded RCs, demonstrate that these represent the whole galaxy kinematics for the region and sample under study and, finally seek a curve capable of accurately describing all the (8) ${\bf 2N}$ coadded RCs\footnote{Let us stress that the acronym URC is used along the following pipeline: a) a large and complete Sample of RCs of galaxies of different luminosities and same Hubble Type b) an ensemble of suitably {\it coadded} velocity data representing the whole kinematics of the above sample, c) an {\it empirical } analytical function that fits well such ensemble and that d) can result the output of a {\it physically} motivated galaxy mass model }

\begin{figure*}[t!]
\centering
\includegraphics[width=13cm]{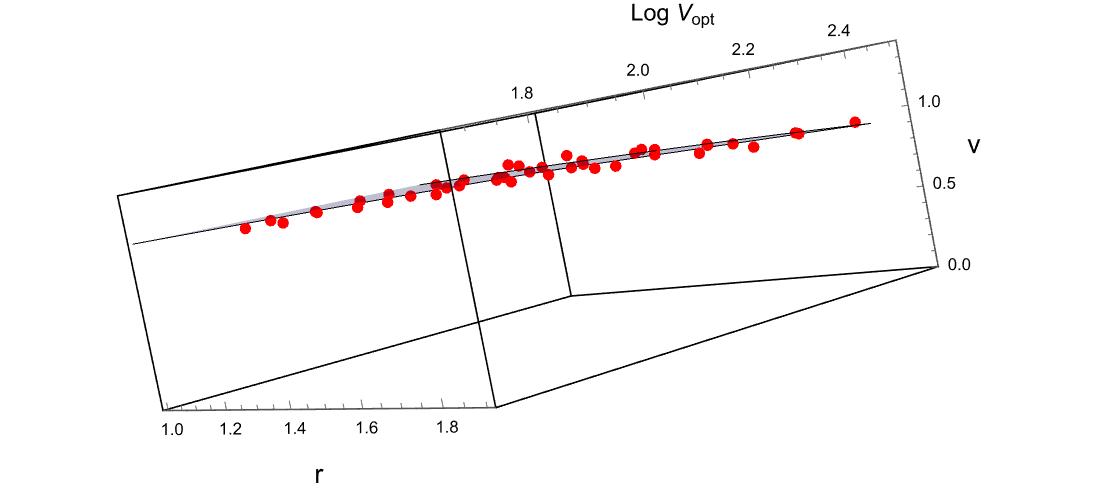}
\includegraphics[width=9cm]{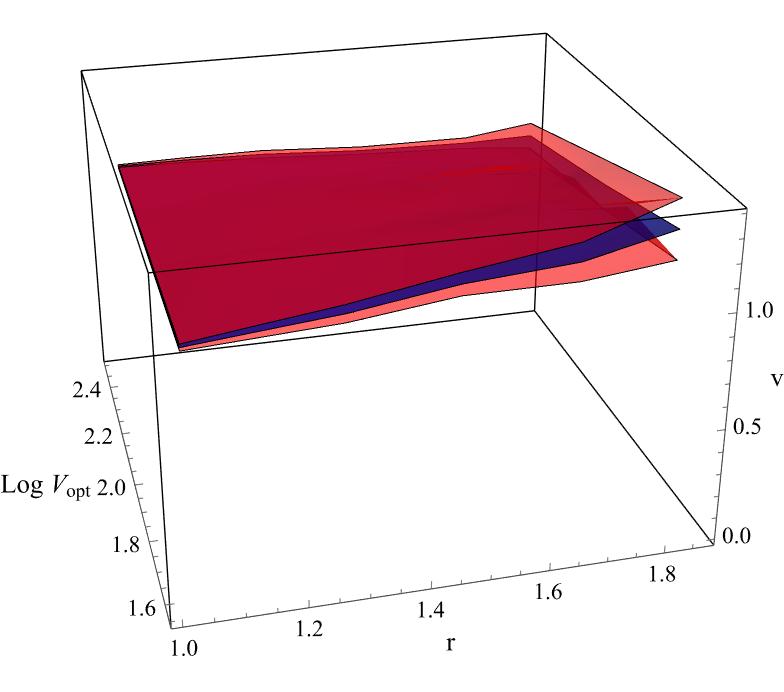}
\includegraphics[width=9cm]{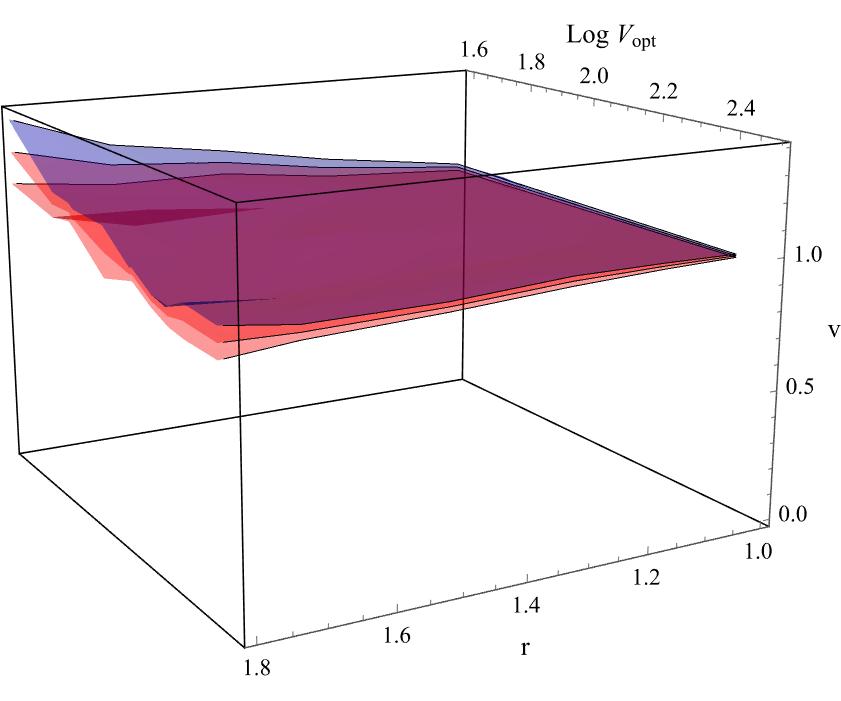}

\caption{The plot visualizes the URC-2opt from the SPARC galaxies. At the top we plot the 48 2N velocity data that form the ensemble of coadded measurements of SPARC from a vantage point illustrating that they belong to an unique line inside the URC concept. The bottom 2 panels represent the 3D smoothed version of $v_{URC-2opt}(r, \log\ V_{opt} )$ manifolds, compared with the the two $\pm \sigma$ surfaces of $v_{URC-vir}(r, \log\ V_{opt} )$ \citep{Salucci:2007tm}, demonstrating the excellent agreement.}
\label{fig:urc_3d}
\end{figure*}

% \rev{Maybe slightly confising as in the bottom panel the blue manifold is outside the two red ones, unlike the second one.}

In \cref{fig:binVopt} we show the resulting ${\bf 2N}$ coadded RC for each of the 8 $\bf VEL$ bins. The resulting surface is smooth and with small statistical irregularities (see \cref{fig:urc_3d}) that creates a ``fundamental" plane relating $v(r)$, $r$ and of $\log \, V_{opt}$: \cref{fig:urc_3d} shows how the 8 coadded RCs form a smooth and well-defined surface. This promotes them to be the backbone of the URC-2opt for S and dIrr galaxies.
% Finally, we work out the empirical form of:
{We determine the analytical form of the empirical URC--2opt by performing a global fit to the ensemble of the 48 coadded double-normalized rotation curve measurements, minimizing a $\chi^2$ function with respect to a low-order expansion in $r$ whose coefficients are allowed to depend on $\log V_{opt}$. The functional form is selected as the lowest-order parametrization capable of reproducing the smooth surface defined by the eight coadded curves. While the linear model provides a simple relationship for the URC, capturing the main global trend of the data, the quadratic fit allows for a gentle curvature matching eventual gradual transitions in rotation curves profile. We avoided to add a cubic term in order to prevent overfitting and non-monotonic behavior in the velocity profiles. We then assume the following analytical expression for the URC-2opt, valid only in the considered region where: $1\leq r \leq 2$.
\begin{equation}
\label{eq:vurc}
v_{URC-2opt}(r,V_{opt})=1+\left[c_0 + c_1\log_{10}V_{opt} + c_2 (\log_{10}V_{opt})^2\right](r-1)
\end{equation}

The best-fit parameters $a$, $b$, and $c$ reported in \cref{tab:urc_fit}. The URC-2opt analytical function accurately reproduces the 48 coadded measurements, with a reduced $\chi^2$ of 2.64 and an r.m.s. scatter of 0.0229 dex, indicating that the URC-2opt provides an excellent fit to the data. Let us notice that $V_{URC-2opt}(R)$ can be obtained by combining \cref{eq:vurc} with \cref{eq:VURC}. In Fig.~\ref{fig:URC_manifolds}, we investigate the URC fit adopting, for the $V_{opt}$ term, a linear and a cubic model in relation with the quadratic one of \cref{eq:vurc}. 
%\rev sandeep: before we excluded the cubic a-priori...
Note that \cref{eq:vurc} defines, in the region $1\leq r\leq 2$ an analytical placeholder fitting function capable of successfully reproducing the coadded data, establishing so the URC-2opt from the SPARC sample. However, other more physically driven functional forms could be adopted, as done in URC-vir. In this work we keep the functional form in \cref{eq:vurc} simple, since we use the URC-2opt emerged from SPARC, mainly to demonstrate its agreement, in the above region, with the URC-PROBES and with the extrapolated URC-vir (see \cref{tab:urc_fit}).

{\renewcommand{\arraystretch}{1.2} 
\setlength{\tabcolsep}{6pt} 
\begin{table*}[t!]
\caption{Best-fit parameters for the Universal Rotation Curve model in \cref{eq:vurc}, obtained through a global fit to the coadded RCs dataset. Uncertainties correspond to $1\sigma$ C.L. inferred from the fitting posterior. Here we only demonstrate the agreement between the two datasets used in this work. For an extended and more complete statistical analysis of other possible URC fitting formulae we refer the reader to App.~\ref{app:urc_fit}.}
\label{tab:urc_fit}
\centering
\begin{tabular}{lcc}
\hline
Parameter & URC-2opt & URC-2opt+URC-zoom \\
\hline
\hline
$c_0$ & 3.1623 $\pm$ 0.3897 & 3.1033 $\pm$ 0.3802 \\
$c_1$ & $-2.4770$ $\pm$ 0.3674 & $-2.4099$ $\pm$ 0.3589 \\
$c_2$ & 0.4724 $\pm$ 0.0861 & 0.4549 $\pm$ 0.0841 \\
\hline
% \multicolumn{3}{c}{ } \\
Number of data points & \multicolumn{1}{c}{48} & \multicolumn{1}{c}{88} \\
Degrees of freedom & \multicolumn{1}{c}{45} & \multicolumn{1}{c}{85} \\
RMS residual & \multicolumn{1}{c}{0.0229} & \multicolumn{1}{c}{0.0222} \\
Reduced $\chi^2$ & \multicolumn{1}{c}{2.64} & \multicolumn{1}{c}{2.16} \\ 
\hline
\end{tabular}
\end{table*}
}

The small uncertainties on the coefficients in \cref{eq:vurc}, as well as the very low root mean square (r.m.s.) scatter of 0.02 dex, demonstrate that the mass distribution information of these galaxies is efficiently captured by a universal function, with $V_{opt}$ and $R_{opt}$ serving as the free observational parameters tagging the individual systems. It is important to emphasize that within the region studied here, both the ensemble of the 48 coadded RC datasets and the corresponding URC given in \cref{eq:vurc} accurately represent the kinematics of the SPARC RC sample. Given the sample's completeness and excellent kinematical resolution, we robustly establish the empirical existence of the URC-2opt (for Spirals and dwarf Irregulars), whether defined analytically or as a tabulated function.

From \cref{fig:binVopt} and \cref{eq:vurc} we see a relation between the value of the URC-2opt slope at $r$ and $V_{opt}$, as generally found in the inner regions of disk systems investigated with individual RCs or with the URC \citep{1996MNRAS.281...27P}. As it occurs within $R_{opt}$, also in the current region under investigation, we seldom find a continuously flat RC. In some case the flattening is likely to occur at $R> 2R_{opt}$, but the RC will not remain flat for long, as in, at these distances the Dark halos densities decrease faster than $R^{-2} $ causing the RC to decline. Comparing \cref{fig:binVopt} with \cref{fig:urc_al} which shows the 105 SPARC RCs in physical units, we realize that their evident ``diversity" arises when we adopt the $(V, R)$ coordinates system and neglect the existing relationship between the structural parameters of the dark and luminous matter that creates a dependence of $v(r)$ on $\log(V_{opt})$. 

Let us compare the URC-2opt (obtained from SPARC) with the corresponding URC-vir of Spirals and dIrrs that extends out to the virial radius \citep{Salucci:2007tm}, but that, however, in the region $R > R_{opt}$ must be considered as an ensemble of curves {\it extrapolated} by means of the mass modelling of the URC-opt which is determined by the spiral's kinematics {\it inside} $R_{opt}$ \citep{1996MNRAS.281...27P}. The normal Spirals in URC-opt come from $\sim 10^5$ individual H$_{\alpha}$ measurements, almost totally inside $R_{opt}$. In \cite{Salucci:2007tm} the URC-opt has been extrapolated out to the virial radius by means of (1) the halo-disk mass decomposition of the latter URC (built out of data located inside $R_{opt}$) (2) the $M_{vir}$ vs $M_{disk}$ relationship obtained via the abundance matching method \citep{Shankar:2006xz}. The result was the URC-vir, the first URC of spirals extended out to $R_{vir}$, but with its coadded curves ensemble extended only out to $R_{opt}$. Moreover, in the SPARC sample, alongside the spirals there are a good number of dwarf Irregulars, with $40 \leq V_{opt} \leq 75 \, \rm {\rm km/s}$. For these systems we have used the extrapolations of the URC established in \citep{2017MNRAS.465.4703K}. 
 
In \cref{fig:binVopt} we show that the URC-2opt is in a very good agreement with the extrapolated URC-vir. Noticeably, in building these two universal curves the RC data employed, their location in the galaxies and the method of analysis are completely independent. This agreement is a strong indication that the URC-2opt, which is directly built from the SPARC RCs, is not only consistent with the URC-vir but also reinforces its validity in the region $R_{opt} \leq R \leq 2\, R_{opt}$, where the URC-vir was previously extrapolated.

{\renewcommand{\arraystretch}{1.2} 
\setlength{\tabcolsep}{6pt} 
\begin{table*}[t!]
\caption{Equivalent to \cref{tab:bin_opticalval} for the PROBES sample.}
\label{tab:bin_opticalval_PROBES}
\centering
\begin{tabular}{@{}cccc@{}}
\toprule
$\langle V_{{opt}}\rangle\,(\mathrm{km\,s^{-1}})$ & $V_{opt, \rm{min}}\,(\mathrm{km\,s^{-1}})$ & $V_{opt, \rm{max}}\,(\mathrm{km\,s^{-1}})$ & $N_{\rm gal}$\\ 
\midrule

% $42.91\pm1.52$&$17.50$&$58.90$&$61$ \\
% \hline
% $68.75\pm0.86$&$58.90$&$77.50$&$49$\\
% \hline
% $83.78\pm0.54$&$77.50$&$90.50$&$51$\\
% \hline
% $111.11\pm0.70$&$90.50$&$129.10$&$264$\\
% \hline
% $146.54\pm0.67$&$129.10$&$165.50$&$274$\\
% \hline
% $185.40\pm0.73$&$165.50$&$209.70$&$277$\\
% \hline
% $226.64\pm1.17$&$209.70$&$250.50$&$98$\\
% \hline
% $285.05\pm4.94$&$250.50$&$423.00$&$44$\\
$42.91$&$17.5$&$58.9$&$61$ \\

$68.75$&$58.9$&$77.5$&$49$\\

$83.78$&$77.5$&$90.5$&$51$\\

$111.11$&$90.5$&$129.1$&$264$\\

$146.54$&$129.1$&$165.5$&$274$\\

$185.4$&$165.5$&$209.7$&$277$\\

$226.64$&$209.7$&$250.5$&$98$\\

$285.05$&$250.5$&$423.0$&$44$\\
\botrule

\end{tabular}

\end{table*}
}

{\renewcommand{\arraystretch}{1.2} 
\setlength{\tabcolsep}{6pt} 
\begin{table*}[t!]
 \caption{Equivalent to \cref{tab:avg-vopt} for the PROBES sample, obtained by combining $5$ {\bf RAD} bins within each of the $8$ optical {\bf VEL} bins. }
 \label{tab:avg-vopt-PROBES}
 \centering
 \vskip 0.5cm
 \begin{minipage}{0.48\textwidth}
 \centering
 \small
 \resizebox{\textwidth}{!}{%
 
 \begin{tabular}{@{}r r r r r r r r@{}}
 \hline
 \multicolumn{7}{c}{$\langle V_{opt} \rangle = 42.91\,\mathrm{km\,s^{-1}}$} \\
 \hline
 $i$ & $N$ & $r_i$ & $v_i$ & $\sigma_i$& $R_i$ & $V_i$ \\
 \hline
 $1$ & $19$ & $0.973$ & $0.980$ & $0.011$ & $3.59$ & $42.06$ \\
 $2$ & $16$ & $1.026$ & $0.993$ & $0.011$ & $3.79$ & $42.61$ \\
 $3$ & $14$ & $1.079$ & $1.017$ & $0.035$ & $3.99$ & $43.62$ \\
 $4$ & $18$ & $1.124$ & $1.074$ & $0.022$ & $4.15$ & $46.10$ \\
 $5$ & $14$ & $1.178$ & $1.100$ & $0.038$ & $4.35$ & $47.20$ \\
 \hline
 \end{tabular}
 }
 \resizebox{\textwidth}{!}{%
 \begin{tabular}{@{}r r r r r r r r@{}}
 \hline
 \multicolumn{7}{c}{$\langle V_{opt} \rangle = 83.78\,\mathrm{km\,s^{-1}}$} \\
 \hline
 $i$ & $N$ & $r_i$ & $v_i$ & $\sigma_i$& $R_i$ & $V_i$ \\
 \hline
 $1$ & $21$ & $0.975$ & $0.963$ & $0.025$ & $7.62$ & $80.68$ \\
 $2$ & $28$ & $1.024$ & $1.026$ & $0.014$ & $8.01$ & $85.96$ \\
 $3$ & $21$ & $1.072$ & $1.028$ & $0.023$ & $8.38$ & $86.15$ \\
 $4$ & $15$ & $1.128$ & $0.986$ & $0.037$ & $8.82$ & $82.65$ \\
 $5$ & $10$ & $1.179$ & $1.068$ & $0.061$ & $9.21$ & $89.46$ \\
 \hline
 \end{tabular}
 }
 \resizebox{\textwidth}{!}{%
 \begin{tabular}{@{}r r r r r r r r@{}}
 \hline
 \multicolumn{7}{c}{$\langle V_{opt} \rangle = 146.54\,\mathrm{km\,s^{-1}}$} \\
 \hline
 $i$ & $N$ & $r_i$ & $v_i$ & $\sigma_i$& $R_i$ & $V_i$ \\
 \hline
 $1$ & $238$ & $0.974$ & $1.002$ & $0.006$ & $10.22$ & $146.78$ \\
 $2$ & $207$ & $1.025$ & $1.012$ & $0.008$ & $10.74$ & $148.32$ \\
 $3$ & $181$ & $1.074$ & $1.008$ & $0.008$ & $11.26$ & $147.74$ \\
 $4$ & $144$ & $1.124$ & $1.005$ & $0.011$ & $11.78$ & $147.29$ \\
 $5$ & $118$ & $1.175$ & $1.030$ & $0.012$ & $12.32$ & $150.91$ \\
 \hline
 \end{tabular}
 }
 \resizebox{\textwidth}{!}{%
 \begin{tabular}{@{}r r r r r r r r@{}}
 \hline
 \multicolumn{7}{c}{$\langle V_{opt} \rangle = 226.64 \,\mathrm{km\,s^{-1}}$} \\
 \hline
 $i$ & $N$ & $r_i$ & $v_i$ & $\sigma_i$& $R_i$ & $V_i$ \\
 \hline
 $1$ & $117$ & $0.976$ & $1.003$ & $0.006$ & $14.95$ & $227.43$ \\
 $2$ & $84$ & $1.026$ & $0.994$ & $0.007$ & $15.71$ & $225.17$ \\
 $3$ & $76$ & $1.075$ & $1.007$ & $0.006$ & $16.47$ & $228.28$ \\
 $4$ & $56$ & $1.125$ & $0.996$ & $0.008$ & $17.23$ & $225.83$ \\
 $5$ & $40$ & $1.175$ & $1.001$ & $0.011$ & $17.99$ & $226.93$ \\
 \hline
 \end{tabular}
 }
 \end{minipage}
 \hfill
 \begin{minipage}{0.48\textwidth}
 \centering
 %\caption{Average $V_{opt}=69.32\,\mathrm{km\,s^{-1}}$}
 
 \small
 \resizebox{\textwidth}{!}{%
 \begin{tabular}{@{}r r r r r r r r@{}}
 \hline
 \multicolumn{7}{c}{$\langle V_{opt} \rangle = 68.75\,\mathrm{km\,s^{-1}}$} \\
 \hline
 $i$ & $N$ & $r_i$ & $v_i$ & $\sigma_i$& $R_i$ & $V_i$ \\
 \hline
 $1$ & $21$ & $0.973$ & $1.010$ & $0.008$ & $6.28$ & $69.47$ \\
 $2$ & $18$ & $1.022$ & $1.018$ & $0.025$ & $6.60$ & $70.01$ \\
 $3$ & $13$ & $1.078$ & $0.979$ & $0.033$ & $6.96$ & $67.32$ \\
 $4$ & $7$ & $1.127$ & $1.050$ & $0.017$ & $7.27$ & $72.22$ \\
 $5$ & $4$ & $1.171$ & $1.049$ & $0.009$ & $7.55$ & $72.11$ \\
 \hline
 \end{tabular}%
 }
 \resizebox{\textwidth}{!}{%
 \begin{tabular}{@{}r r r r r r r r@{}}
 \hline
 \multicolumn{7}{c}{$\langle V_{opt} \rangle = 111.11\,\mathrm{km\,s^{-1}}$} \\
 \hline
 $i$ & $N$ & $r_i$ & $v_i$ & $\sigma_i$& $R_i$ & $V_i$ \\
 \hline
 $1$ & $198$ & $0.975$ & $1.014$ & $0.008$ & $8.81$ & $112.70$ \\
 $2$ & $174$ & $1.023$ & $1.003$ & $0.008$ & $9.25$ & $111.47$ \\
 $3$ & $133$ & $1.074$ & $1.012$ & $0.010$ & $9.71$ & $112.45$ \\
 $4$ & $112$ & $1.122$ & $1.039$ & $0.013$ & $10.15$ & $115.45$ \\
 $5$ & $105$ & $1.176$ & $1.042$ & $0.016$ & $10.63$ & $115.80$ \\
 \hline
 \end{tabular}
 }
 \resizebox{\textwidth}{!}{%
 \begin{tabular}{@{}r r r r r r r r@{}}
 \hline
 \multicolumn{7}{c}{$\langle V_{opt} \rangle = 185.40\,\mathrm{km\,s^{-1}}$} \\
 \hline
 $i$ & $N$ & $r_i$ & $v_i$ & $\sigma_i$& $R_i$ & $V_i$ \\
 \hline
 $1$ & $336$ & $0.974$ & $1.001$ & $0.004$ & $12.67$ & $185.52$ \\
 $2$ & $303$ & $1.026$ & $1.007$ & $0.005$ & $13.33$ & $186.64$ \\
 $3$ & $267$ & $1.075$ & $1.007$ & $0.005$ & $13.98$ & $186.77$ \\
 $4$ & $212$ & $1.124$ & $0.997$ & $0.007$ & $14.61$ & $184.82$ \\
 $5$ & $203$ & $1.174$ & $1.003$ & $0.008$ & $15.26$ & $185.88$ \\
 \hline
 \end{tabular}
 }
 \resizebox{\textwidth}{!}{%
 \begin{tabular}{@{}r r r r r r r r@{}}
 \hline
 \multicolumn{7}{c}{$\langle V_{opt} \rangle = 285.05\,\mathrm{km\,s^{-1}}$} \\
 \hline
 $i$ & $N$ & $r_i$ & $v_i$ & $\sigma_i$& $R_i$ & $V_i$ \\
 \hline
 $1$ & $40$ & $0.972$ & $1.013$ & $0.011$ & $17.39$ & $288.68$ \\
 $2$ & $30$ & $1.026$ & $0.994$ & $0.012$ & $18.34$ & $283.47$ \\
 $3$ & $27$ & $1.074$ & $0.975$ & $0.014$ & $19.21$ & $278.06$ \\
 $4$ & $22$ & $1.124$ & $1.017$ & $0.015$ & $20.09$ & $289.96$ \\
 $5$ & $15$ & $1.173$ & $0.934$ & $0.027$ & $20.98$ & $266.12$ \\
 \hline
 \end{tabular}
 }
\end{minipage}
\end{table*}
}

As a consequence of these results it is relevant to investigate, with high-resolution {\it optical} RCs, the URC profile at the edge of the stellar disks \textit{e.g.}, the region defined by : $ 0.95 \ R_{opt} \leq R \ \leq 1.2 \ R_{opt}$. This region includes the reference radius $R_{opt}$ and it works as a buffer between the region probed by optical RCs and that probed by HI RCs. Furthermore, it includes the edge of the stellar disk and witnesses the rise of the HI disk as the main component of the baryonic matter density. 
To perform the task we exploit the optical rotation curves very recently provided by the PROBES sample and analyzed in \cite{2022ApJS..262...33S}. PROBES (Photometry and Rotation Curves Observations from Extragalactic Surveys) is the largest available catalog of RCs of disk systems which, in addition, are provided with surface brightness measures that yield us the values of $R_{opt}$. We use, in the above region, a total of 3677 independent measurements from 1118 galaxies \cref{tab:avg-vopt-PROBES}, more than the double of those used, in the same region, in building the URC-1996. that provide us with the URC-zoom, \textit{i.e.}, 5 coadded measures at radii around $R_{opt}$ and covering a region $0.25\, R_{opt}$ wide. All the stages of the building of this URC are very similar to those set in \cite{1996MNRAS.281...27P} and discussed in \Cref{sec:urc2p0}. Furthermore, we have adopted a subdivision of the individual RCs very similar to that used for the URC-2opt pipeline (see \cref{tab:bin_opticalval}). We determine five values of the ${\bf 2N}$ coadded rotation velocity relative to 5 radial bins of equal width $\Delta r = 0.05$ covering the buffer region $0.95 \le r \le 1.20$ (see \cref{tab:bin_opticalval_PROBES,tab:avg-vopt-PROBES}).

We find, an excellent agreement between all the three URCs. As a consequence, the URC-zoom reinforces the URC-96 at the crucial radius of the stellar disk edge and both reinforce the URC-2opt by showing in this region it is the same profile and hence, underlying physics. In the second column of \cref{tab:urc_fit} we also report the best-fit parameters obtained by fitting the URC-2opt to the 48 coadded measurements including the 40 additional measurements from the PROBES sample. The resulting best-fit parameters are consistent within $1\sigma$ with those obtained when using only the SPARC data, and the fit quality is slightly improved, with a reduced $\chi^2$ of 2.16 and an r.m.s. scatter of 0.0222. While the inclusion of the PROBES data does not significantly alter the best-fit parameters of the URC-2opt, it does provide a slightly better fit to the combined dataset. The agreement between the URC-zoom derived from PROBES and the SPARC-based URC-2opt is particularly significant given the substantial differences in tracers, spatial resolution, and survey systematics. That these independent datasets recover the same normalized kinematic behavior in the narrow buffer region around $R_{opt}$ providing a strong consistency check on the universality of the rotation curve and rules out the possibility that the URC arises from survey-specific biases footnote{In regard to using the SPARC rotation curves (RCs) to rigorously test specific dark matter (DM) halo density profiles or modifications of gravity, it is important to note that approximately one third of the sample's RCs should be excluded, as they are either not usable or do not provide reliable results for mass modeling.}. 

% \TS{I think these two paragraphs above are new compared to the previous version, so they should be turned in red color.}
Given the results of this work it is worth noticing that the presence, in the range $1\leq r\leq 2$, of the URCs investigated/built here of several cases of distinctly rising or declining profiles is inconsistent with the predictions of both the MOND and the NFW DM halo + Freeman stellar disk scenarios which envisage rather flat RCs in this radial range \citep{1996ApJ...462..563N}. 

It is important to highlight that to adopt blindly a density profile without an \textit{a priori} direct reference to observational data is likely to be unsuccessful. In fact, DM halos have a quite complicated "general profile" that very likely can be determined only by reverse engineering the available observations.

\section{Radial Tully-Fisher relation}
\label{sec:RTF}
We now seek to explore the SPARC sample in the context of the relationships identified in spiral and dwarf irregular galaxies, specifically those involving $V(r)$, $r$, and $Mag$. As a starting point, consider the well-established Tully–Fisher (TF) relation \citep{1977A&A....54..661T}, which links the absolute magnitude $Mag$ of a spiral galaxy (in a given photometric band) to the Full Width Half Maximum (FWHM) of the 21 cm emission line of neutral hydrogen. The FWHM serves as a reliable proxy for the galaxy’s maximum rotation velocity ($V_{max}$), a physically meaningful characteristic that has led to an improved form of the TF relation where $V_{max}$ replaces the FWHM:
\begin{equation}
 \label{eq:tf}
 Mag=A~\log V_{max}+B
\end{equation}
where $\rm{A}$ and $\rm{B}$ are constants, with $A \simeq 8 \pm 1$ depending on the chosen luminosity band and galaxy sample. A significant challenge of the TF relation concerns the choice of rotational velocity used as reference. This is because $V(R)$ varies with radius $R$ in ways that depend strongly on a galaxy’s structural properties. For instance, in galaxies with low luminosity and low $V_{opt}$, the rotation curve typically rises steadily, reaching its maximum well outside the edge of the stellar disc, that is, $R_{max} > R_{opt}$. Conversely, in more luminous galaxies, the maximum velocity is found in the very central regions, with $R_{max}\simeq (1/5) R_{opt}$. Moreover, providing a clear physical interpretation of the TF relation in \cref{eq:tf} is challenging. While the TF correlation undoubtedly stems from the application of the virial theorem to rotating galaxies, it remains unclear how its parameters $A$ and $B$, as well as the intrinsic scatter, are affected by variations in key galactic properties—such as the dark matter content and distribution, the stellar disk mass-to-light ratio, or the central surface brightness. Although the tightness and simplicity of \cref{eq:tf} may result from intricate interrelations among multiple galactic structural parameters, it is not possible to disentangle or "reverse-engineer" these dependencies. For example, if a given galaxy deviates significantly from the TF relation, we cannot determine from this alone whether the discrepancy arises from an incorrect distance estimate or from atypical dark matter content or distribution.
 
\citep{2007MNRAS.377..507Y} introduced the RTF relation as a refinement of the classical TF relation. Note that the RTF relation is not a simple extension of the TF relation, but rather a fundamentally different approach to correlating galaxy luminosity with kinematic properties. Also, the RTF relation is not a single relation but an ensemble of relations, each corresponding to a specific normalized radius $r$ on the stellar disk. Moreover, designed to address its limitations and reduce the associated intrinsic scatter which is often attributed to variations in galaxy properties that are not accounted for by the simple correlation between luminosity and maximum rotation velocity. By incorporating the radial dependence of the rotation curve, the RTF relation aims to capture more of the underlying physics that governs galaxy dynamics, thereby reducing this scatter. 

While the classical TF relation correlates galaxy luminosity with a reference velocity measured at a fixed physical radius, the RTF relation further incorporates the specific location on the stellar disk, expressed by the normalized coordinate $r \equiv R/R_{opt}$, at which the velocity is measured. This explicitly accounts for the radial dependence of the kinematic–luminosity relation within galaxies. This new reference velocity is intrinsically different from that determined by the velocity measured at a fixed reference radius used in the TF relationship.(\textit{e.g.}, the radius where it has the maximum value). 

We have that, at fixed normalized radii $r$, tight relations between the galaxy luminosity and the velocity $V(r)$ at these radii emerge. The number of these relations depends on he maximum number of independent velocity measurements in the radial range $0.2<r<1 $, that are available for all the galaxies of a sample. In detail, the sample's galaxies enjoy an ensemble of relations, named the RTF relationships, whose number (from 5 to 8) is equal to the above maximum\footnote{Given a sample there is, of course, a trade off between the number of relations we aim to set up and the number of objects participating in each of them.}.
We refer the reader to our previous work, where we demonstrated that this relationship arises directly from the underlying distributions of dark and luminous matter in disk galaxies. In detail, we divide the galactic region $R_{in}\leq R\leq R_{out}$ with available disk kinematics in $k$ equally spaced radial bins, each with a suitable number of rotation velocity data. For simplicity, lets set: $R_{in}=0$, $R_{out}=R_{opt}$. The bins are of size $R_{opt}/k$ and each of them is centred at
$R_c(m)=(m -1/2)\, R_{opt}/k$, with: $1\leq m\leq k$ \footnote{The final inferences do not depend on the choice of the bin characteristics, provided that $R_c(m)\propto R_{opt}$.}. Then, for each $m$-named bin one gets $V (R_c(m))$, the average velocity of all the RC measurements in the bin $m$ and investigates the existence of $k$ independent TF-like relations, each one of the form:
\begin{equation}
\label{eq:radrtf}
 Mag =a_m +b_m \,\log \, V (R_c(m)),
\end{equation} 
each characterized by a root mean square (r.m.s.) scatter $\sigma_m$. The existence of the RTF relationship was firmly demonstrated by \citep{2007MNRAS.377..507Y}, who analyzed three independent samples of spiral galaxies. In each case, a family of TF-like relations, following \cref{eq:radrtf} was found to emerge.

\begin{figure*}[t!]
 \centering

 % --- Row 1 ---

 \centering
 \includegraphics[width=\linewidth]{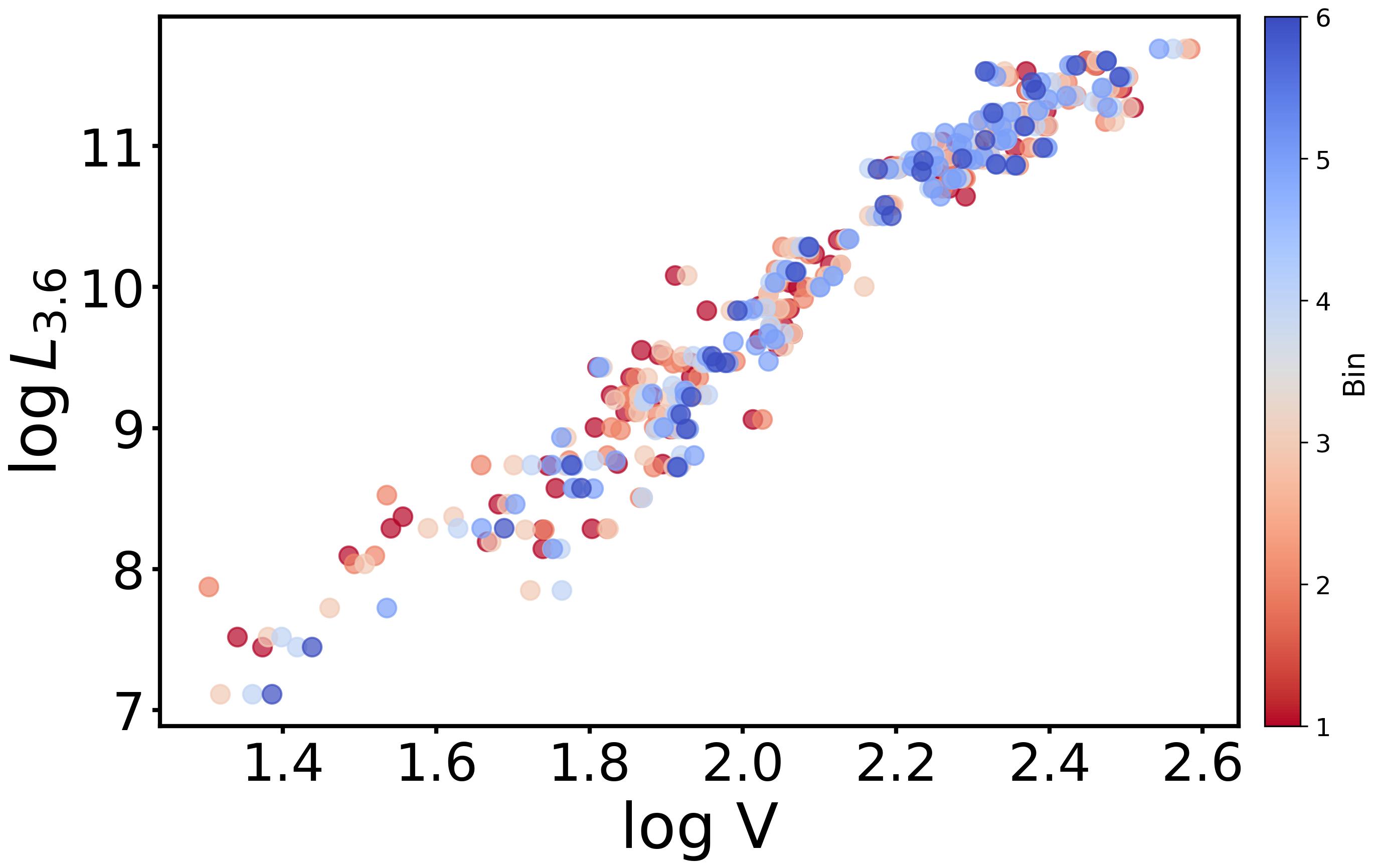}
 %\caption{Bin 1},height=0.57\textheight,clip

 \caption{Radial RTF relation for the SPARC galaxies. The left plot represents the RTF relation at radial distance of $j~R_{opt}$ where $j$ is (1.05, 1.15, 1.3, 1.5, 1.7, 1.9) for the 105 selected SPARC galaxies.}
 \label{fig:rtf}
\end{figure*}

These relations exhibit significantly smaller scatter than their corresponding classical TF relations for the same samples, and their slopes $b_m$ increase markedly with normalized radial distance $m/k\, R_{opt}$ \citep{2007MNRAS.377..507Y} (see \cref{fig:rtf}). The RTF encapsulates crucial information about the mass distribution in spirals because it fundamentally reflects the intertwined distribution of dark and luminous matter. Unlike the classical TF relation, the RTF inherently accommodates galaxies that share the same luminosity and/or $V_{max}$ but differ in their dark matter fractions or other structural parameters—factors that often undermine the applicability of the standard TF relation. Moreover, the physically motivated nature of the RTF enables the identification of outlier galaxies or anomalous observational data with greater clarity. In this context, we analyze the SPARC sample using \cref{eq:radrtf}, specifically searching for the existence of these independent TF-like relations at radii $R > R_{opt}$, extending the investigation beyond the optical RC region.

To obtain a reliable proxy for the stellar disk mass $M_D$, it is standard to use red or infrared band magnitudes for $Mag$ (e.g., previous studies have used $R$ and $I_C$ bands). In the SPARC sample, IR 3.6~$\mu$m band magnitudes (luminosities) are available for all objects and serve as an excellent tracer of the stellar disk mass, following $M_D = \mathrm{const} \times L_{3.6}$ \citep{2014AJ....148...77M, Meidt:2014mqa}. For the present analysis, we focus on the region extending from the edge of the stellar disk, $R = 3.2\,R_D$, out to $R = 6\,R_D$, which marks the outermost radius where a sufficient number of 21~cm rotation curves is currently available.

We utilize 129 rotation curves\footnote{Twenty SPARC RCs lack sufficient data to be included in the analysis. They are lesser than those excluded in building the URC-2opt, since here we accept also the case of RCs with just one (bin) measurement in the region we consider} from the SPARC sample to investigate the existence of the RTF relationship in the region previously discussed. This region lies well {\it beyond} the zone where the relation was originally identified $(0 - R_{opt})$, while it is of comparable radial width $(R_{opt} - 2R_{opt})$. Specifically, in addition to $L_{\mathrm{3.6\,\mu m}}$, we define the reference velocities following this scheme: for each RC, we compute an average velocity within radial bins centered at $j\,R_{opt}$, with $j = \{1.05, 1.15, 1.3, 1.5, 1.7, 1.9\}$, and bin widths of $\{0.1\,R_{opt}, 0.1\,R_{opt}, 0.2\,R_{opt}, 0.2\,R_{opt}, 0.2\,R_{opt}, 0.2\,R_{opt}\}$, respectively. This approach ensures a similar number of velocity measurements in each radial bin, providing a balanced sampling across the investigated region.

We then fit the $\sim$ 600 velocity data obtained from the SPARC sample with $6$ independent linear relations of the form:
\begin{equation}\label{eq:rtf}
 \log L_{3.6} = b_j\, \log V_j + a_j
\end{equation}
that define, for our sample, the family of RTF relations. $a_j$, $b_j$ are the free parameters constrained from the observations and $V_j$ is the velocity measured at the reference radius $j=R/R_{opt}$. It is worth noticing that the last radial bin in \cite{2007MNRAS.377..507Y}, using data from \cite{1996MNRAS.281...27P} coincides with the first radial bin in this work. {For each radial bin $j$, the parameters $(a_j,b_j)$ are determined via a linear least-squares regression in logarithmic space. We minimize the $\chi^2$ defined using the observational uncertainties on $\log L_{3.6}$ and $\log V_j$, allowing for an intrinsic scatter term $\sigma_{\rm int}$, the log-likelihood is written as,
\begin{equation}
\label{eq:loglike_rtf}
\ln \mathcal{L}_j(a_j,b_j,\sigma_{\rm int}) =
-\frac{1}{2}\sum_{i=1}^{N_j}
\left[
\frac{\left(\log L_{3.6,i} - b_j \log V_{j,i} - a_j\right)^2}
{\sigma_{i,j}^2 + \sigma_{\rm int}^2}
+ \ln\left(2\pi (\sigma_{i,j}^2 + \sigma_{\rm int}^2)\right)
\right].
\end{equation}

The uncertainties on the fitted parameters are then derived from the covariance matrix of the fit, while the intrinsic scatter of each relation is quantified by the r.m.s. of the residuals about the best-fitting line. } 

In \cref{fig:rtf}, we show the ensemble of RTF relations (see also \cref{fig:rtf_fit} for relations plotted individually) for the SPARC galaxies and \cref{tab:rtf_fit} reports the corresponding constraints on the RTF parameters. We realize the existence of TF-like relations at the fixed $j \ R_{opt}$ radii. Importantly, we recognize an increase of the slope $a(j) $ of these relations as function of radius $j \ R_{opt}$. Remarkably, the SPARC $a(j)$ values smoothly connect at $R_{opt}$ to those of the RTF relations established in the inner region $(0.2\,R_{opt} - R_{opt})$. This continuity confirms that the RTF relationship identified at smaller radii extends out to $R_{opt}$, highlighting the robustness and universality of the RTF across both luminous and dark matter dominated regions. In other words, we find that at smaller radii, within the current dark matter dominated outer regions, the same trend observed in the luminous mass dominated optical region persists \citep{2007MNRAS.377..507Y}. Specifically, the slope of the RTF relation increases with radius, indicating that the correlation between luminosity and rotation velocity becomes stronger as we move outward from the galaxy center. This trend is consistent with the expectation that dark matter plays a more significant role in the outer regions of galaxies, influencing the kinematics in a way that enhances the luminosity-velocity correlation.

 The region inside $R_{opt}$ is well covered by the URC-vir, URC-dIrr and URC-LSB, with the latter in good agreement with the SPARC individual RCs (amounting to $\sim 1/8$ those concurring in the URCs). In the region $R_{opt}$-$2 R_{opt}$, which is insufficiently {\it directly} covered by the above mentioned URCs, the SPARC sample results crucial. In fact, there, it has sufficient kinematical measures to demonstrate the existence of the URC emerging in agreement with the URC built inside $R_{opt}$ and then extrapolated to $2\, R_{opt}$. In practice, as far as the mass distribution in disk in Spirals all the information of the SPARC sample is compacted in \cref{eq:vurc}-\cref{eq:rtf} and \cref{tab:avg-vopt}-\cref{tab:urc_fit} and \cref{tab:rtf_fit}.

To facilitate a direct comparison between the results of the two samples, we examine the SPARC RTF relation for the bin centered at $2.2\,R_D$. This radius is particularly noteworthy as it corresponds to the minimum scatter in the RTF and exhibits a significantly shallower slope in the TF-like relation compared to more external bins. To further test the universality of the RTF, we analyze the relation at $j=0.6$ for the SPARC sample. This radius lies within the optical region and is where the tightest correlation of the RTF ensemble is observed. We find that the parameters of the SPARC RTF relation at this location show very good agreement with those from the Yegorova $\&$ Salucci \cite{2007MNRAS.377..507Y} sample (see \cref{fig:rtf2}).

\begin{figure*}[ht!]
 \centering

 \centering
 \includegraphics[width=0.9\textwidth]{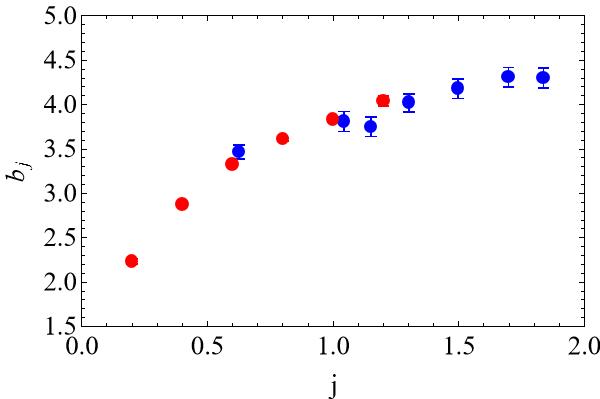}
 % #16cm

 \caption{The plot represents the slopes $b_j$ of the RTF \cref{eq:rtf} built from the HI RC at the various radial distances $j\equiv R/R_{opt}$ in the region $R_{opt}-6~R_D$ and indicated in the text. {\it Blue points} Also shown the slopes of the RTF obtained by optical RC out to $R_{opt}$ in \citep{2007MNRAS.377..507Y}.}
 \label{fig:rtf2}
\end{figure*}
 
\section{Conclusion}
\label{sec:conc}
 
In disk systems, the URC, built by co-adding a very large number of RCs, embodies the structural properties of the dark and luminous mass distributions, including their systematic relations. On the other hand, a sample like SPARC of $\sim100$ objects with very extended high resolution {\it individual} RCs, allows one to investigate the aforementioned aspects by probing every RC separately and out to larger distances from their centers. In this work, we have shown that these two approaches are complementary. In the region $R_{opt} - 2R_{opt}$, the Universal Rotation Curves determined so far for disk systems and listed in Table 3 are not derived from direct coadded RC measurements ( as for $R\leq R_{opt}$ range, but rather from the {\it extrapolation} of the URC constructed within $R_{opt}$, which serves to establish the various galaxy mass models for galaxy. In detail, in this outer region, the SPARC coadded RC {\it measurements} result in good agreement with the (extrapolated) URC-vir and, around $R_{opt}$, with a zoomed URC out of 1118 Probes galaxies. 

This supports the notion of universality in the disk system RCs: the slopes and amplitudes of the rotation curves in the inner ($R < R_{opt}$) regions can reliably predict those in the outer ($R_{opt} < R < 2R_{opt}$) regions. In short, in this work, by means of the SPARC sample we have directly built the URC (for normal spirals) out to a radius twice as far than those of the previous URCs, listed in Table 3. 
 
In the context of the most critical investigations on the dark and luminous matter in galaxies, the Universal Rotation Curve, either in the analytical or in the coadded data formulation, is not constrained by any specific theoretical model and works as a portal towards the New Physics that the DM phenomenon entails.

%{This profile, which can be used to test the validity of various theoretical predictions. For instance, if a particular theoretical model predicts a specific rotation curve shape, we can directly compare this prediction against the empirically derived URC. If the model's predicted curve significantly deviates from the URC, it may indicate that the model does not accurately capture the underlying physics of galaxy rotation curves. Conversely, if the model's predictions align well with the URC, it would lend support to the model's validity. This observationally grounded approach allows for a more robust and direct testing of theoretical models against empirical data.}

Finally, investigating the DM phenomenon with a large sample of individual RCs requires, for each analysis, firstly to reconstruct the intrinsic systematic properties of the RCs out of the raw data, and then investigate these properties in the adopted line of thought. By contrast, the URC allows us to bypass the initial step: the overall structural properties of disk galaxy RCs have already been systematically encoded once for all in a model independent way, either through a analytical expression or via just about 1000 "measurements" of the coadded RCs.

\backmatter

% \bmhead{Supplementary information}

\section*{Acknowledgements}

PS acknowledges Iniziativa specifica QGSKY.

% Acknowledgements are not compulsory. Where included they should be brief. Grant or contribution numbers may be acknowledged.

\section*{Declarations}

\textbf{Funding:} This research was supported by progetto PRIN 2022 (2022EEZRR3) - PE9 - ``The quest for the understanding of the Dark matter phenomenon'', funded by the European Union - Next Generation EU - PNRR Investimento M4.C2.1.1 (CUP: G53C24000830006; funding agency: MUR).

\textbf{Conflict of interest/Competing interests:} The authors declare that they have no conflict of interest.

\textbf{Ethics approval and consent to participate:} Not applicable.

\textbf{Consent for publication:} Not applicable.

\textbf{Data availability:} The datasets analyzed during the current study are publicly available at \url{http://astroweb.cwru.edu/SPARC} (SPARC sample). Additional data or results generated during the study are available from the corresponding author upon reasonable request.

\textbf{Materials availability:} Not applicable.

\textbf{Code availability:} The code used to process and analyze the data in this study is available from the corresponding author upon reasonable request.

\textbf{Author contributions:} EB and PS have contributed to the conception, methodology and preformed the main analysis with the SPARC dataset and produced the primary results in the manuscript. TS and SH have contributed to the analysis of the PROBES dataset and to the comparison between of URC-2opt, obtained from SPARC and PROBES dataset. All authors contributed to the interpretation of the results and to the preparation of the manuscript. All authors read and approved the final manuscript.

% \noindent
% If any of the sections are not relevant to your manuscript, please include the heading and write `Not applicable' for that section. 

\begin{appendices}

\section{List of Abbreviations}
\label{sec:abb}
This section compiles the list of symbols and terminologies used throughout the paper which are compiled in tabular format in \cref{tab:abb}
{\renewcommand{\arraystretch}{1.3} 
\setlength{\tabcolsep}{6pt} 
\begin{table*}[h!]
 \caption{The table explains the various terminology used throughout the paper. }
\label{tab:abb}
\centering
\begin{tabular}{|c|c|}
\hline
Abbreviation & Name \\ 
\hline
\hline
$\bf O $ region & optical region: from the center out to $R_{opt}$\\
\hline
$\bf HI$ region & from $R_{opt} $ out to $2 R_{opt} $\\
\hline
$R_D$ & length scale of the Freeman exponential thin disk\\
\hline
$R_{opt}$ & $3.2 \ R_{disk}$. Reference size of the stellar disk in spirals \\
\hline
RC & Rotation Curve\\
\hline
$V(R)$ & circular velocity in km/s\\
\hline
$V_{opt}$ & $V(R_{opt})$ \\
\hline
$2N$ & double normalized quantity at a radius $r=R/R_{opt}$ and with $v=V(R)/V_{opt} $\\
\hline
$2N$ RC & $v(r)$ \\
\hline
URC & generic Universal Rotation Curve\\
\hline
TF & Tully Fisher relationship\\
\hline
RTF & Radial Tully Fisher. An ensemble of new and specific TF-like relations\\
\hline
${\bf R,I_C,3.6\, \mu m }$ & systems of magnitudes \\
\hline
$L_{3.6}$ & The Spitzer 3.6 $\mu m$ band luminosity \\ 
\hline
\textbf{VEL} bin & Binning based on $V_{opt}$ (optical circular velocity) \\ 
\hline
\textbf{RAD} bin & Binning based on radius or normalized radius $r=R/R_{opt}$ \\ 
\hline
DM & Dark Matter \\ 
\hline
dIrr & dwarf Irregular galaxy \\ 
\hline
LSB & Low Surface Brightness galaxy \\ 
\hline
NFW & Navarro-Frenk-White density profile \\ 
\hline
URC-0 & First claimed Universal Rotation Curve \\
\hline
URC-opt & Universal Rotation Curve obtained for the optical region \\
\hline
URC-vir & Universal Rotation Curve extrapolated to the virial radius \\ 
\hline
URC-2opt & Universal Rotation Curve obtained for the region $R_{opt}$--$2R_{opt}$ \\ 
\hline
URC-zoom & Universal Rotation Curve in the region $0.95~R_{opt}$--$1.2~R_{opt}$ \\ 
\hline
SPARC & Spitzer Photometry and Accurate Rotation Curves galaxy sample \\ 
\hline
PROBES & PROBES rotation curve sample (gathering data from various sources) \\

\hline

\end{tabular}

\end{table*}}

\section{RTF fit parameters}
\label{app:RTF_fit}

In this section, we briefly present the best-fit parameters of the RTF relations obtained from the SPARC sample, as well as the corresponding fits to the data. The parameters include the slope ($b_j$), intercept ($a_j$), and intrinsic scatter ($\sigma_{\rm int}$) for each of the TF-like relations at different radial bins. The results are compiled in \cref{tab:rtf_fit} and visually represented in \cref{fig:rtf_fit}.

{\renewcommand{\arraystretch}{1.2} 
\setlength{\tabcolsep}{6pt} 
\begin{table*}[t!]
\caption{The best-fit parameters of the RTF: the slope ($b_j$), the intercept ($a_j$) and intrinsic scatter $\sigma_{\rm int}$ for each of its TF-like relations. }
\label{tab:rtf_fit}
\centering
 
% \vskip 0.5truecm
\begin{tabular}{@{}llll@{}}
\toprule

Bin (${ j}$) & $b_{ j}$ & $a_{ j}$ & $\sigma_{ int, j}$ \\ 
\midrule
$1$ & $3.81 \pm 0.11$ & $2.10 \pm 0.24$ & $0.27$ \\

$2$ & $3.75 \pm 0.11$ & $2.24 \pm 0.23$ & $0.26$ \\

$3$ & $4.02 \pm 0.10$ & $1.64 \pm 0.21$ & $0.28$ \\

$4$ & $4.18 \pm 0.11$ & $1.31 \pm 0.23$ & $0.26$ \\

$5$ & $4.31 \pm 0.11$ & $1.03 \pm 0.24$ & $0.23$ \\

$6$ & $4.30 \pm 0.15$ & $1.06 \pm 0.31$ & $0.23$ \\
\botrule
\end{tabular}

\end{table*}
}

 % The uncertainties in these parameters are represented by $\sigma_a$ and $\sigma_b$, respectively.

\begin{figure*}[t!]
 \centering

 % --- Row 1 ---

 \centering
 \includegraphics[width=\linewidth]{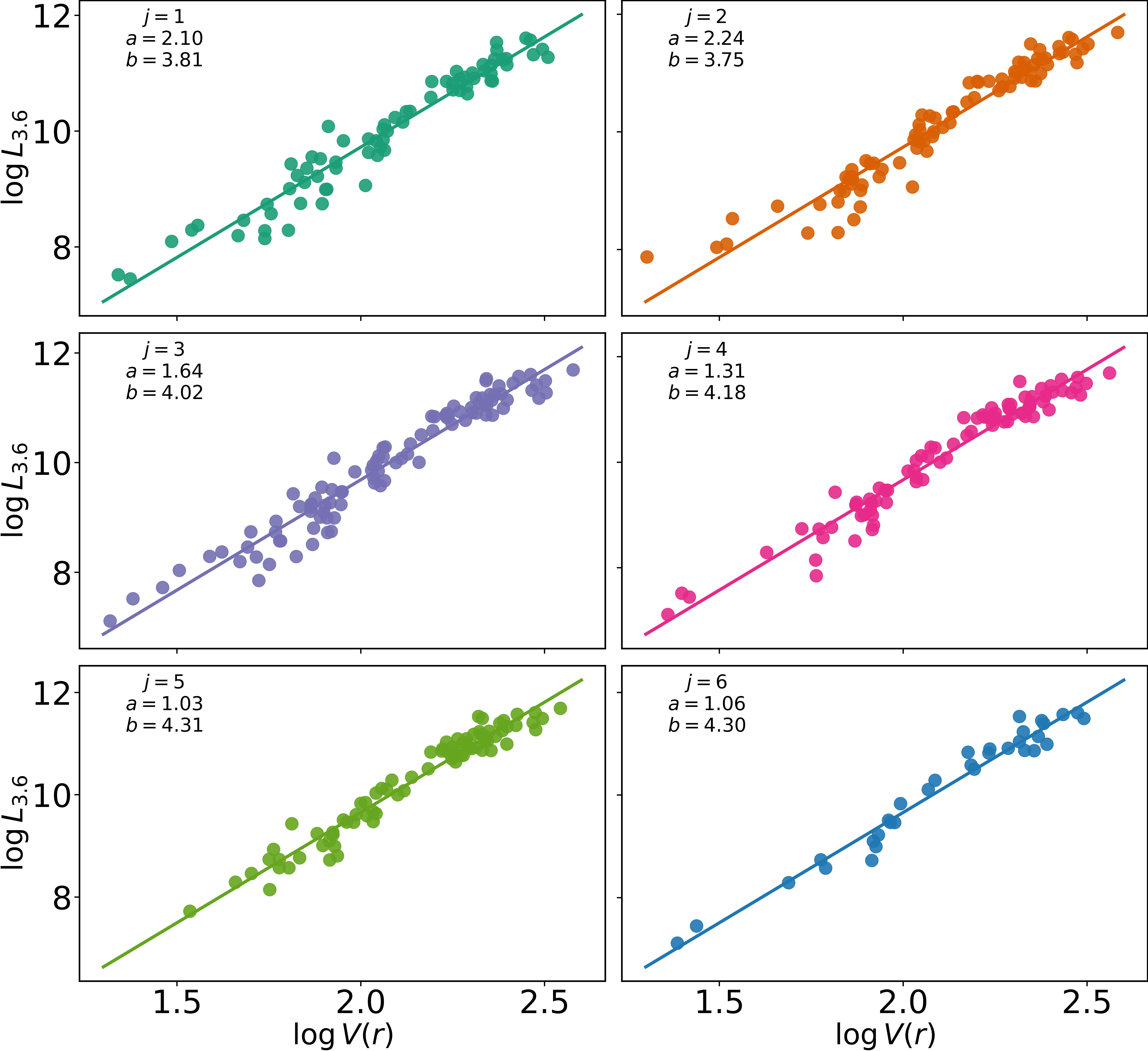}
 %\caption{Bin 1},height=0.57\textheight,clip

 \caption{Radial RTF relation for the SPARC galaxies. From the top-left panel, each figure represents the RTF for individual bin as compiled in Table \ref{tab:rtf_fit}. The individual RTF obtained from SPARC are fitted to Eq.\ref{eq:rtf} with slope (b) and intercept (a) as specified in individual plot as well as in Table \ref{tab:rtf_fit}. }
 \label{fig:rtf_fit}
\end{figure*}

\section{On the URC-2opt fitting formula}
\label{app:urc_fit}
In this section, we discuss briefly the fitting function to the URC velocity. While in the main text we have utilized only the quadratic model to demonstrate the point, here we do extended analysis with linear and cubic formulae. As elaborated in the main text the purpose of \cref{tab:urc_fit} is to show that the URC-2opt can be well fitted by a simple quadratic formula in $\log_{10}(V_{opt})$ and that there is a good agreement with the URC-zoom derived from the PROBES sample. In \cref{fig:URC_manifolds} we show the fit of the URC model to the SPARC co-added data, for the linear, quadratic and cubic in $\log_{10}(V_{opt})$ models.
\begin{figure}
 \centering
 \includegraphics[width=1.0\linewidth]{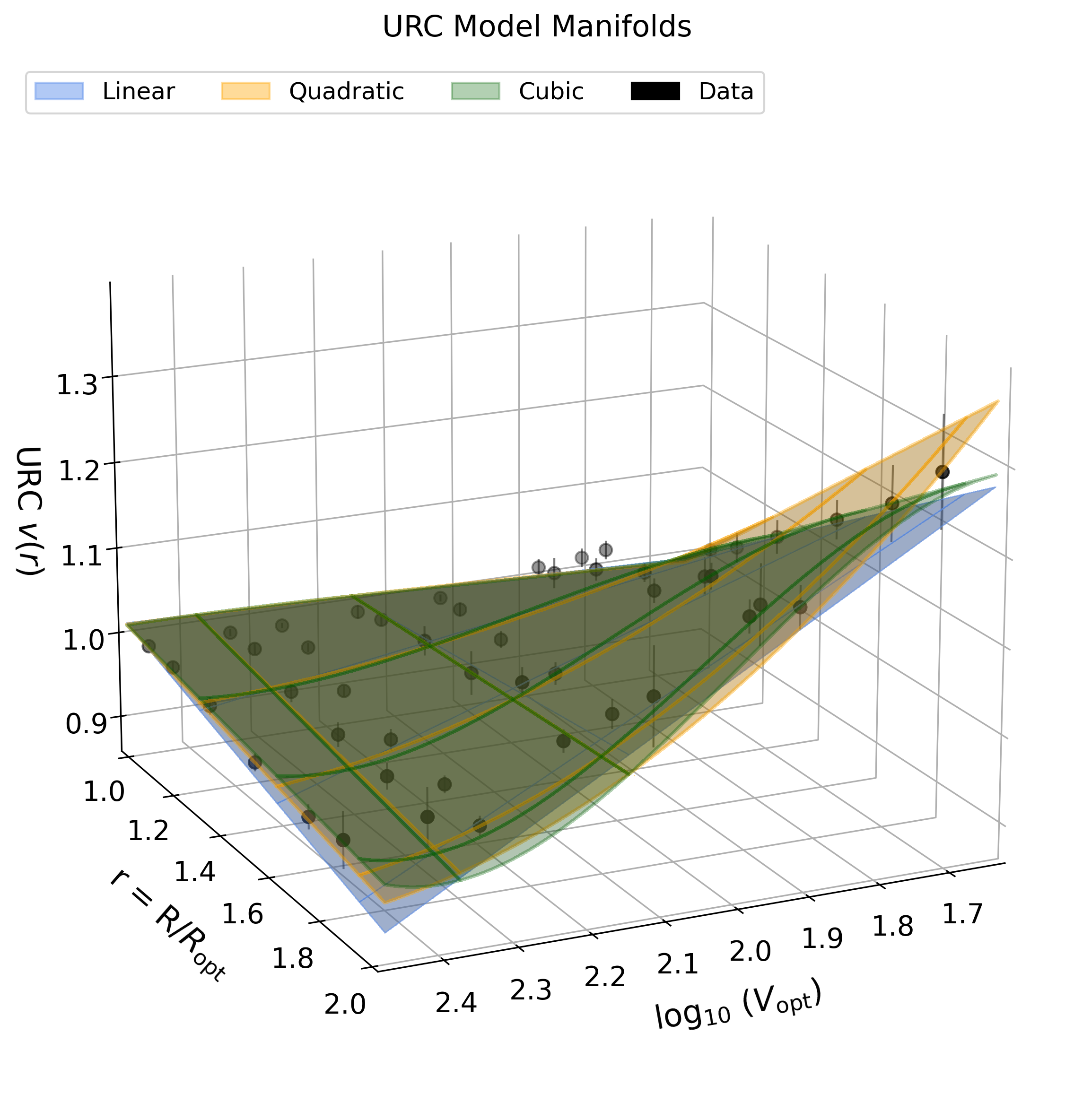}
 \caption{We show the fit of URC models to the SPARC co-added data, for the linear (blue), quadratic (orange, \cref{eq:vurc} in the main text) and cubic (green) in $\log_{10}(V_{opt})$ model dependence alongside the data points.}
 \label{fig:URC_manifolds}
\end{figure}

To assess the performance of the URC models we refine the likelihood utilized to fit the data by including an global intrinsic scatter parameter which is necessary to anticipate better the observed scatter in the data. The log-likelihood is written as,
\begin{equation}
\label{eq:loglike_urc}
\ln \mathcal{L}(\mathbf{p},\sigma_{\rm int}) =
-\frac{1}{2}\sum_{i=1}^{N}
\left[
\frac{\left(v_i - v_{\rm URC}(r_i, V_{opt}; \mathbf{p})\right)^2}
{\sigma_i^2 + \sigma_{\rm int}^2}
+ \ln\left(2\pi (\sigma_i^2 + \sigma_{\rm int}^2)\right)
\right],
\end{equation}
where $\mathbf{p}\equiv \{c_n\}$ are the parameters of the URC model written as $(v_{\rm URC}(r, V_{opt}; \mathbf{p})-1)/(r-1) = \sum_{n=0}^N c_n(\log_{10}V_{opt})^n$, $v_i$ and $\sigma_i$ are the observed velocity and its uncertainty for the $i^{\rm th}$ data point, and $\sigma_{\rm int}$ is the global intrinsic scatter of the model. The best-fit parameters are obtained by maximizing this log-likelihood function. The uncertainties in the fitted parameters can be derived from the covariance matrix of the fit, while the intrinsic scatter can be quantified by the r.m.s. of the residuals about the best-fitting curve. 

We then estimate the Akaike Information Criterion (AIC) \cite{Akaike1974} for each model to compare their relative performance. The AIC is given by, $\mathrm{AIC} = 2k - 2\ln(\hat{\mathcal{L}})$, where $k$ is the number of parameters in the model and $\hat{\mathcal{L}}$ is the maximum value of the likelihood function. We find that the linear model is disfavored compared to the cubic model with a $\Delta \mathrm{AIC} \sim 5.0$\footnote{Models with $\Delta \mathrm{AIC} \leq 2$ can be practically considered indistinguishable, while those with $4 \leq \Delta \mathrm{AIC} \leq 7$ are less preferable, and finally $\Delta \mathrm{AIC} > 10$ indicate that the data strongly disfavor the model \cite{BurnhamAnderson2002, BurnhamAnderson2004}.}, while the quadratic model is indistinguishable w.r.t the cubic model with a $\Delta \mathrm{AIC} \sim 0.2$, essentially validating that the lowest-order model that can fit the data is the quadratic one. The corresponding parameters constraints and the model-selection criteria are presented in \cref{tab:model_comparison}. Finally, in addition to the polynomial functions we also test a few extensions,

\begin{align}
 v_{URC-\mathrm{PL}}(r) &= 1 + \alpha \left[1 - \left(\frac{V_{\mathrm{opt}}}{V_0}\right)^{\beta}\right](r-1),
 \\
 v_{URC-\tanh}(r) &= 1 + \alpha \tanh\!\left[\beta \,\log_{10}\!\left(\frac{V_{\mathrm{opt}}}{V_0}\right)\right](r-1).
\end{align}

which are both disfavored against the quadratic model with $\Delta \mathrm{AIC} \simeq 7.0$.

{\renewcommand{\arraystretch}{1.2}
\setlength{\tabcolsep}{6pt} 
\begin{table*}[t!]
\caption{Constraints on parameters for the linear, quadratic, and cubic models using the SPARC co-added data. The parameters $c_n$ corresponding to the coefficients of the respective models, while $\sigma_{\rm int}$ represents the global intrinsic scatter of the data for a given model. Uncertainties correspond to $1\sigma$ errors. Note, in contrast to the \cref{tab:urc_fit}, here we also include the intrinsic scatter parameter to correctly perform the model-selection. Which in turn, provides the differences to the fit parameters of the quadratic model w.r.t the values presented therein.}
\label{tab:model_comparison}
\centering
\begin{tabular}{lccc}
\hline
Parameter & Linear & Quadratic & Cubic \\
\hline
\hline
$c_0$ & 1.0391 $\pm$ 0.0700 & 2.4926 $\pm$ 0.5000 & 1.0884 $\pm$ 4.0000 \\
$c_1$ & $-0.4628$ $\pm$ 0.0300 & $-1.8542$ $\pm$ 0.4000 & 0.2829 $\pm$ 5.0000 \\
$c_2$ & --- & 0.3290 $\pm$ 0.1000 & $-0.7416$ $\pm$ 3.0000 \\
$c_3$ & --- & --- & 0.1767 $\pm$ 0.4000 \\
$\sigma_{\rm int}$ & 0.0154 $\pm$ 0.0070 & 0.0128 $\pm$ 0.2000 & 0.0120 $\pm$ 0.2000 \\
\hline
RMS residual & \multicolumn{1}{c}{0.0209} & \multicolumn{1}{c}{0.0201} & \multicolumn{1}{c}{0.0199} \\
Reduced $\chi^2$ & \multicolumn{1}{c}{1.013} & \multicolumn{1}{c}{1.1} & \multicolumn{1}{c}{1.148} \\
% $\chi^2$ & \multicolumn{1}{c}{$-233.3$} & \multicolumn{1}{c}{$-240.1$} & \multicolumn{1}{c}{$-242.3$} \\
$\Delta$AIC & \multicolumn{1}{c}{$5.0$} & \multicolumn{1}{c}{$0.2$} & \multicolumn{1}{c}{$0$} \\
\hline
\end{tabular}
\end{table*}
}

\end{appendices}

\bibliography{sn-bibliography}% common bib file

@ARTICLE{2016AJ....152..157L,
       author = {{Lelli}, Federico and {McGaugh}, Stacy S. and {Schombert}, James M.},
        title = "{SPARC: Mass Models for 175 Disk Galaxies with Spitzer Photometry and Accurate Rotation Curves}",
      journal = {\aj},
         year = 2016,
        month = dec,
       volume = {152},
       number = {6},
          eid = {157},
        pages = {157},
          doi = {10.3847/0004-6256/152/6/157},
archivePrefix = {arXiv},
       eprint = {1606.09251},
 primaryClass = {astro-ph.GA},
       adsurl = {https://ui.adsabs.harvard.edu/abs/2016AJ....152..157L}
}

@ARTICLE{1991ApJ...368...60P,
       author = {{Persic}, Massimo and {Salucci}, Paolo},
        title = "{The Universal Galaxy Rotation Curve}",
      journal = {\apj},
         year = 1991,
        month = feb,
       volume = {368},
        pages = {60},
          doi = {10.1086/169670},
       adsurl = {https://ui.adsabs.harvard.edu/abs/1991ApJ...368...60P}
}

@ARTICLE{1996MNRAS.281...27P,
       author = {{Persic}, Massimo and {Salucci}, Paolo and {Stel}, Fulvio},
        title = "{The universal rotation curve of spiral galaxies {\textemdash} I. The dark matter connection}",
      journal = {\mnras},
         year = 1996,
        month = jul,
       volume = {281},
       number = {1},
        pages = {27-47},
          doi = {10.1093/mnras/278.1.27},
archivePrefix = {arXiv},
       eprint = {astro-ph/9506004},
 primaryClass = {astro-ph},
       adsurl = {https://ui.adsabs.harvard.edu/abs/1996MNRAS.281...27P}
}

@ARTICLE{2017MNRAS.465.4703K,
       author = {{Karukes}, E.~V. and {Salucci}, P.},
        title = "{The universal rotation curve of dwarf disc galaxies}",
      journal = {\mnras},
         year = 2017,
        month = mar,
       volume = {465},
       number = {4},
        pages = {4703-4722},
          doi = {10.1093/mnras/stw3055},
archivePrefix = {arXiv},
       eprint = {1609.06903},
 primaryClass = {astro-ph.GA},
       adsurl = {https://ui.adsabs.harvard.edu/abs/2017MNRAS.465.4703K}
}

@ARTICLE{1977A&A....54..661T,
       author = {{Tully}, R.~B. and {Fisher}, J.~R.},
        title = "{A new method of determining distances to galaxies.}",
      journal = {\aap},
         year = 1977,
        month = feb,
       volume = {54},
        pages = {661-673},
       adsurl = {https://ui.adsabs.harvard.edu/abs/1977A&A....54..661T}
}

@ARTICLE{2007MNRAS.377..507Y,
       author = {{Yegorova}, Irina A. and {Salucci}, Paolo},
        title = "{The radial Tully-Fisher relation for spiral galaxies - I}",
      journal = {\mnras},
         year = 2007,
        month = may,
       volume = {377},
       number = {2},
        pages = {507-515},
          doi = {10.1111/j.1365-2966.2007.11637.x},
archivePrefix = {arXiv},
       eprint = {astro-ph/0612434},
 primaryClass = {astro-ph},
       adsurl = {https://ui.adsabs.harvard.edu/abs/2007MNRAS.377..507Y}
}

@ARTICLE{2001AJ....122.2381M,
       author = {{McGaugh}, Stacy S. and {Rubin}, Vera C. and {de Blok}, W.~J.~G.},
        title = "{High-Resolution Rotation Curves of Low Surface Brightness Galaxies. I. Data}",
      journal = {\aj},
         year = 2001,
        month = nov,
       volume = {122},
       number = {5},
        pages = {2381-2395},
          doi = {10.1086/323448},
archivePrefix = {arXiv},
       eprint = {astro-ph/0107326},
 primaryClass = {astro-ph},
       adsurl = {https://ui.adsabs.harvard.edu/abs/2001AJ....122.2381M}
}

@ARTICLE{2009A&A...493..871S,
       author = {{Swaters}, R.~A. and {Sancisi}, R. and {van Albada}, T.~S. and {van der Hulst}, J.~M.},
        title = "{The rotation curves shapes of late-type dwarf galaxies}",
      journal = {\aap},
         year = 2009,
        month = jan,
       volume = {493},
       number = {3},
        pages = {871-892},
          doi = {10.1051/0004-6361:200810516},
archivePrefix = {arXiv},
       eprint = {0901.4222},
 primaryClass = {astro-ph.CO},
       adsurl = {https://ui.adsabs.harvard.edu/abs/2009A&A...493..871S}
}

@ARTICLE{2019MNRAS.490.5451D,
       author = {{Di Paolo}, Chiara and {Salucci}, Paolo and {Erkurt}, Adnan},
        title = "{The universal rotation curve of low surface brightness galaxies - IV. The interrelation between dark and luminous matter}",
      journal = {\mnras},
         year = 2019,
        month = dec,
       volume = {490},
       number = {4},
        pages = {5451-5477},
          doi = {10.1093/mnras/stz2700},
archivePrefix = {arXiv},
       eprint = {1805.07165},
 primaryClass = {astro-ph.GA},
       adsurl = {https://ui.adsabs.harvard.edu/abs/2019MNRAS.490.5451D}
}

@ARTICLE{2016PhRvL.117t1101M,
       author = {{McGaugh}, Stacy S. and {Lelli}, Federico and {Schombert}, James M.},
        title = "{Radial Acceleration Relation in Rotationally Supported Galaxies}",
      journal = {\prl},
         year = 2016,
        month = nov,
       volume = {117},
       number = {20},
          eid = {201101},
        pages = {201101},
          doi = {10.1103/PhysRevLett.117.201101},
archivePrefix = {arXiv},
       eprint = {1609.05917},
 primaryClass = {astro-ph.GA},
       adsurl = {https://ui.adsabs.harvard.edu/abs/2016PhRvL.117t1101M}
}

@ARTICLE{2019A&ARv..27....2S,
       author = {{Salucci}, Paolo},
        title = "{The distribution of dark matter in galaxies}",
      journal = {\aapr},
         year = 2019,
        month = feb,
       volume = {27},
       number = {1},
          eid = {2},
        pages = {2},
          doi = {10.1007/s00159-018-0113-1},
archivePrefix = {arXiv},
       eprint = {1811.08843},
 primaryClass = {astro-ph.GA},
       adsurl = {https://ui.adsabs.harvard.edu/abs/2019A&ARv..27....2S}
}

@ARTICLE{2020ApJS..247...31L,
       author = {{Li}, Pengfei and {Lelli}, Federico and {McGaugh}, Stacy and {Schombert}, James},
        title = "{A Comprehensive Catalog of Dark Matter Halo Models for SPARC Galaxies}",
      journal = {\apjs},
         year = 2020,
        month = mar,
       volume = {247},
       number = {1},
          eid = {31},
        pages = {31},
          doi = {10.3847/1538-4365/ab700e},
archivePrefix = {arXiv},
       eprint = {2001.10538},
 primaryClass = {astro-ph.GA},
       adsurl = {https://ui.adsabs.harvard.edu/abs/2020ApJS..247...31L}
}

@ARTICLE{2017MNRAS.470.2410R,
       author = {{Rodrigues}, Davi C. and {del Popolo}, Antonino and {Marra}, Valerio and {de Oliveira}, Paulo L.~C.},
        title = "{Evidence against cuspy dark matter haloes in large galaxies}",
      journal = {\mnras},
         year = 2017,
        month = sep,
       volume = {470},
       number = {2},
        pages = {2410-2426},
          doi = {10.1093/mnras/stx1384},
archivePrefix = {arXiv},
       eprint = {1701.02698},
 primaryClass = {astro-ph.GA},
       adsurl = {https://ui.adsabs.harvard.edu/abs/2017MNRAS.470.2410R}
}

@ARTICLE{2018JCAP...08..012D,
       author = {{de Almeida}, {\'A}lefe and {Amendola}, Luca and {Niro}, Viviana},
        title = "{Galaxy rotation curves in modified gravity models}",
      journal = {\jcap},
         year = 2018,
        month = aug,
       volume = {2018},
       number = {8},
          eid = {012},
        pages = {012},
          doi = {10.1088/1475-7516/2018/08/012},
archivePrefix = {arXiv},
       eprint = {1805.11067},
 primaryClass = {astro-ph.GA},
       adsurl = {https://ui.adsabs.harvard.edu/abs/2018JCAP...08..012D}
}

@ARTICLE{2024arXiv240110202K,
       author = {{Khelashvili}, Mariia and {Rudakovskyi}, Anton and {Hossenfelder}, Sabine},
        title = "{SPARC galaxies prefer Dark Matter over MOND}",
      journal = {arXiv e-prints},
         year = 2024,
        month = jan,
          eid = {arXiv:2401.10202},
        pages = {arXiv:2401.10202},
          doi = {10.48550/arXiv.2401.10202},
archivePrefix = {arXiv},
       eprint = {2401.10202},
 primaryClass = {astro-ph.CO},
       adsurl = {https://ui.adsabs.harvard.edu/abs/2024arXiv240110202K}
}

@ARTICLE{1996ApJ...462..563N,
       author = {{Navarro}, Julio F. and {Frenk}, Carlos S. and {White}, Simon D.~M.},
        title = "{The Structure of Cold Dark Matter Halos}",
      journal = {\apj},
         year = 1996,
        month = may,
       volume = {462},
        pages = {563},
          doi = {10.1086/177173},
archivePrefix = {arXiv},
       eprint = {astro-ph/9508025},
 primaryClass = {astro-ph},
       adsurl = {https://ui.adsabs.harvard.edu/abs/1996ApJ...462..563N}
}

@ARTICLE{2022ApJS..262...33S,
       author = {{Stone}, Connor and {Courteau}, St{\'e}phane and {Arora}, Nikhil and {Frosst}, Matthew and {Jarrett}, Thomas H.},
        title = "{PROBES. I. A Compendium of Deep Rotation Curves and Matched Multiband Photometry}",
      journal = {\apjs},
         year = 2022,
        month = sep,
       volume = {262},
       number = {1},
          eid = {33},
        pages = {33},
          doi = {10.3847/1538-4365/ac83ad},
archivePrefix = {arXiv},
       eprint = {2209.09912},
 primaryClass = {astro-ph.GA},
       adsurl = {https://ui.adsabs.harvard.edu/abs/2022ApJS..262...33S}
}

@ARTICLE{2018ApJ...858...62K,
       author = {{Karachentsev}, Igor D. and {Makarova}, Lidia N. and {Tully}, R. Brent and {Rizzi}, Luca and {Shaya}, Edward J.},
        title = "{TRGB Distances to Galaxies in Front of the Virgo Cluster}",
      journal = {\apj},
         year = 2018,
        month = may,
       volume = {858},
       number = {1},
          eid = {62},
        pages = {62},
          doi = {10.3847/1538-4357/aabaf1},
archivePrefix = {arXiv},
       eprint = {1804.00469},
 primaryClass = {astro-ph.GA},
       adsurl = {https://ui.adsabs.harvard.edu/abs/2018ApJ...858...62K}
}

@ARTICLE{2019ApJ...880...63M,
       author = {{McQuinn}, Kristen. B.~W. and {Boyer}, Martha and {Skillman}, Evan D. and {Dolphin}, Andrew E.},
        title = "{Using the Tip of the Red Giant Branch As a Distance Indicator in the Near Infrared}",
      journal = {\apj},
         year = 2019,
        month = jul,
       volume = {880},
       number = {1},
          eid = {63},
        pages = {63},
          doi = {10.3847/1538-4357/ab2627},
archivePrefix = {arXiv},
       eprint = {1904.01571},
 primaryClass = {astro-ph.GA},
       adsurl = {https://ui.adsabs.harvard.edu/abs/2019ApJ...880...63M}
}

@article{Fouque:2003tm,
    author = "Fouque, Pascal and Storm, Jesper and Gieren, Wolfgang",
    title = "{Calibration of the distance scale from Cepheids}",
    eprint = "astro-ph/0301291",
    archivePrefix = "arXiv",
    doi = "10.1007/978-3-540-39882-0_2",
    journal = "Lect. Notes Phys.",
    volume = "635",
    pages = "21--44",
    year = "2003"
}

@ARTICLE{1996ApJS..107..239T,
       author = {{Taylor}, A.~R. and {Goss}, W.~M. and {Coleman}, P.~H. and {van Leeuwen}, J. and {Wallace}, B.~J.},
        title = "{A Westerbork Synthesis Radio Telescope 327 MHz Survey of the Galactic Plane}",
      journal = {\apjs},
         year = 1996,
        month = nov,
       volume = {107},
        pages = {239},
          doi = {10.1086/192363},
       adsurl = {https://ui.adsabs.harvard.edu/abs/1996ApJS..107..239T}
}

@ARTICLE{2013MNRAS.432.1294P,
       author = {{Petrov}, Leonid and {Mahony}, Elizabeth K. and {Edwards}, Philip G. and {Sadler}, Elaine M. and {Schinzel}, Frank K. and {McConnell}, David},
        title = "{Australia Telescope Compact Array observations of Fermi unassociated sources}",
      journal = {\mnras},
         year = 2013,
        month = jun,
       volume = {432},
       number = {2},
        pages = {1294-1302},
          doi = {10.1093/mnras/stt550},
archivePrefix = {arXiv},
       eprint = {1301.2386},
 primaryClass = {astro-ph.CO},
       adsurl = {https://ui.adsabs.harvard.edu/abs/2013MNRAS.432.1294P}
}

@article{Freeman:1970mx,
    author = "Freeman, K. C.",
    title = "{On the disks of spiral and SO Galaxies}",
    doi = "10.1086/150474",
    journal = "Astrophys. J.",
    volume = "160",
    pages = "811",
    year = "1970"
}

@article{Mo:1997vb,
    author = "Mo, H. J. and Mao, Shude and White, Simon D. M.",
    title = "{The Formation of galactic disks}",
    eprint = "astro-ph/9707093",
    archivePrefix = "arXiv",
    doi = "10.1046/j.1365-8711.1998.01227.x",
    journal = "Mon. Not. Roy. Astron. Soc.",
    volume = "295",
    pages = "319",
    year = "1998"
}

@article{Kravtsov2012THESR,
  title={THE SIZE–VIRIAL RADIUS RELATION OF GALAXIES},
  author={Andrey V. Kravtsov},
  journal={The Astrophysical Journal Letters},
  year={2012},
  volume={764},
  url={https://api.semanticscholar.org/CorpusID:119204308}
}

@article{Salucci:2007tm,
    author = "Salucci, Paolo and Lapi, A. and Tonini, C. and Gentile, G. and Yegorova, I. and Klein, U.",
    title = "{The Universal Rotation Curve of Spiral Galaxies. 2. The Dark Matter Distribution out to the Virial Radius}",
    eprint = "astro-ph/0703115",
    archivePrefix = "arXiv",
    doi = "10.1111/j.1365-2966.2007.11696.x",
    journal = "Mon. Not. Roy. Astron. Soc.",
    volume = "378",
    pages = "41--47",
    year = "2007"
}

@ARTICLE{2014AJ....148...77M,
       author = {{McGaugh}, Stacy S. and {Schombert}, James M.},
        title = "{Color-Mass-to-light-ratio Relations for Disk Galaxies}",
      journal = {\aj},
         year = 2014,
        month = nov,
       volume = {148},
       number = {5},
          eid = {77},
        pages = {77},
          doi = {10.1088/0004-6256/148/5/77},
archivePrefix = {arXiv},
       eprint = {1407.1839},
 primaryClass = {astro-ph.GA},
       adsurl = {https://ui.adsabs.harvard.edu/abs/2014AJ....148...77M}
}

@article{Meidt:2014mqa,
    author = "Meidt, Sharon E. and others",
    title = "{Reconstructing the stellar mass distributions of galaxies using S$^4$G IRAC 3.6 and 4.5 $\mu$m images: II. The conversion from light to mass}",
    eprint = "1402.5210",
    archivePrefix = "arXiv",
    primaryClass = "astro-ph.GA",
    doi = "10.1088/0004-637X/788/2/144",
    journal = "Astrophys. J.",
    volume = "788",
    pages = "144",
    year = "2014"
}

@ARTICLE{Saluccirev21,
       author = {{Salucci}, Paolo and others},
        title = "{Einstein, Planck and Vera Rubin: relevant encounters between the Cosmological and the Quantum Worlds}",
      journal = {Frontiers in Physics},
         year = 2021,
        month = feb,
       volume = {8},
          eid = {579},
        pages = {579},
          doi = {10.3389/fphy.2020.603190},
archivePrefix = {arXiv},
       eprint = {2011.09278},
 primaryClass = {gr-qc},
       adsurl = {https://ui.adsabs.harvard.edu/abs/2021FrP.....8..579S}
}

@ARTICLE{Dipaolorev,
       author = {{Salucci}, Paolo and {di Paolo}, Chiara},
        title = "{Fundamental Properties of the Dark and the Luminous Matter from the Low Surface Brightness Discs}",
      journal = {Universe},
         year = 2021,
        month = sep,
       volume = {7},
       number = {9},
          eid = {344},
        pages = {344},
          doi = {10.3390/universe7090344},
       adsurl = {https://ui.adsabs.harvard.edu/abs/2021Univ....7..344S}
}

@ARTICLE{2018FoPh,
       author = {{Salucci}, Paolo},
        title = "{Dark Matter in Galaxies: Evidences and Challenges}",
      journal = {Foundations of Physics},
         year = 2018,
        month = oct,
       volume = {48},
       number = {10},
        pages = {1517-1537},
          doi = {10.1007/s10701-018-0209-5},
archivePrefix = {arXiv},
       eprint = {1807.08541},
 primaryClass = {astro-ph.GA},
       adsurl = {https://ui.adsabs.harvard.edu/abs/2018FoPh...48.1517S}
}

@article{Shankar:2006xz,
    author = "Shankar, Francesco and Lapi, A. and Salucci, P. and De Zotti, G. and Danese, L.",
    title = "{New relationships between galaxy properties and host halo mass, and the role of feedbacks in galaxy formation}",
    eprint = "astro-ph/0601577",
    archivePrefix = "arXiv",
    doi = "10.1086/502794",
    journal = "Astrophys. J.",
    volume = "643",
    pages = "14--25",
    year = "2006"
}

@ARTICLE{LSD,
       author = {{Lapi}, A. and {Salucci}, P. and {Danese}, L.},
        title = "{Precision Scaling Relations for Disk Galaxies in the Local Universe}",
      journal = {\apj},
         year = 2018,
        month = may,
       volume = {859},
       number = {1},
          eid = {2},
        pages = {2},
          doi = {10.3847/1538-4357/aabf35},
archivePrefix = {arXiv},
       eprint = {1804.06086},
 primaryClass = {astro-ph.GA},
       adsurl = {https://ui.adsabs.harvard.edu/abs/2018ApJ...859....2L}
}

@article{Akaike1974,
  author  = {Akaike, Hirotugu},
  title   = {A New Look at the Statistical Model Identification},
  journal = {IEEE Transactions on Automatic Control},
  volume  = {19},
  number  = {6},
  pages   = {716--723},
  year    = {1974}
}

@book{BurnhamAnderson2002, 
  author    = {Burnham, Kenneth P. and Anderson, David R.},
  title     = {Model Selection and Multimodel Inference: A Practical Information-Theoretic Approach},
  edition   = {2},
  publisher = {Springer},
  year      = {2002}
}

@article{BurnhamAnderson2004, 
  author  = {Burnham, Kenneth P. and Anderson, David R.},
  title   = {Multimodel Inference: Understanding AIC and BIC in Model Selection},
  journal = {Sociological Methods \& Research},
  volume  = {33},
  number  = {2},
  pages   = {261--304},
  year    = {2004}
}

@ARTICLE{Noord,
       author = {{Noordermeer}, E. and {van der Hulst}, J.~M. and {Sancisi}, R. and {Swaters}, R.~S. and {van Albada}, T.~S.},
        title = "{The mass distribution in early-type disc galaxies: declining rotation curves and correlations with optical properties}",
      journal = {\mnras},
         year = 2007,
        month = apr,
       volume = {376},
       number = {4},
        pages = {1513-1546},
          doi = {10.1111/j.1365-2966.2007.11533.x},
archivePrefix = {arXiv},
       eprint = {astro-ph/0701731},
 primaryClass = {astro-ph},
       adsurl = {https://ui.adsabs.harvard.edu/abs/2007MNRAS.376.1513N}
}

@ARTICLE{Catinella, 
       author = {{Catinella}, Barbara and {Giovanelli}, Riccardo and {Haynes}, Martha P.},
        title = "{Template Rotation Curves for Disk Galaxies}",
      journal = {\apj},
         year = 2006,
        month = apr,
       volume = {640},
       number = {2},
        pages = {751-761},
          doi = {10.1086/500171},
archivePrefix = {arXiv},
       eprint = {astro-ph/0512051},
 primaryClass = {astro-ph},
       adsurl = {https://ui.adsabs.harvard.edu/abs/2006ApJ...640..751C}
}

@ARTICLE{Lopez,
       author = {{L{\'o}pez Fune}, E.},
        title = "{Empirical velocity profiles for galactic rotation curves}",
      journal = {\mnras},
         year = 2018,
        month = apr,
       volume = {475},
       number = {2},
        pages = {2132-2163},
          doi = {10.1093/mnras/stx3245},
archivePrefix = {arXiv},
       eprint = {1712.03880},
 primaryClass = {astro-ph.GA},
       adsurl = {https://ui.adsabs.harvard.edu/abs/2018MNRAS.475.2132L}
}

@INPROCEEDINGS{Rhee, 
       author = {{Rhee}, M.-H.},
        title = "{Synthetic Rotation Curves of Spiral Galaxies: a PCA Approach}",
    booktitle = {Dark and Visible Matter in Galaxies and Cosmological Implications},
         year = 1997,
       editor = {{Persic}, Massimo and {Salucci}, Paolo},
       series = {Astronomical Society of the Pacific Conference Series},
       volume = {117},
        month = jan,
        pages = {90},
       adsurl = {https://ui.adsabs.harvard.edu/abs/1997ASPC..117...90R}
}
%% if required, the content of.bbl file can be included here once bbl is generated

\end{document}